 \newif\ifpreprint
   \preprintfalse      
   \preprinttrue       

 \newif\ifpdf

\ifpreprint
        \documentclass[twoside,preprint2]{aastex}
        \ifx\pdfoutput\undefined\pdffalse\else\pdfoutput=1\pdftrue\fi
        \usepackage{amsmath,graphics}

        \ifpdf
           \usepackage{hyperref}
        \fi
\else
        \documentclass[12pt,preprint]{aastex}
        \ifx\pdfoutput\undefined\pdffalse\else\pdfoutput=1\pdftrue\fi
        \usepackage{amsmath}
\fi

\newcommand{\etal}{{et al.}}

\ifpreprint
   \newcommand{\ODcaption}[1] {\caption[]{\footnotesize #1}}
\else
   \newcommand{\ODcaption}[1] {\figcaption{#1}}
\fi

\newcommand{\B}[1]      {{\boldsymbol{#1}}}
\newcommand{\cp}[1]     {c_{\pi#1}}
\renewcommand{\sp}[1]   {s_{\pi#1}}
\renewcommand{\o}[1]    {{\overline{#1}}}
\newcommand{\ave}[1]    {\left\langle{#1}\right\rangle}
\newcommand{\avet}[1]   {\langle{#1}\rangle}

\newcommand{\kms}       {\mbox{${\rm km}\,{\rm s}^{-1}$}}
\newcommand{\kpc}       {\mbox{${\rm kpc}$}}
\newcommand{\kmskpc}    {\mbox{${\rm km}\,{\rm s}^{-1}\,{\rm kpc}^{-1}$}}
\newcommand{\masyr}     {\mbox{${\rm mas}\,{\rm yr}^{-1}$}}

\newcommand{\Sec}[1]    {\S\ref{sec:#1}}
\newcommand{\Secsto}[2] {\S\S\ref{sec:#1} to \ref{sec:#2}}
\newcommand{\eq}[1]     {(\ref{eq:#1})}
\newcommand{\eqn}[1]    {equation~(\ref{eq:#1})}
\newcommand{\eqs}[1]    {equations~(\ref{eq:#1})}
\newcommand{\Tab}[1]    {Table~\ref{tab:#1}}
\newcommand{\Fig}[1]    {Figure~\ref{fig:#1}}
\newcommand{\fig}[1]    {Fig.~\ref{fig:#1}}

\slugcomment{Submitted for publication by the Astrophysical Journal}

\begin{document}

\ifpreprint \thispagestyle{empty} \fi
\title{The Oort Constants Measured from Proper Motions}

\author{Rob P.~Olling\altaffilmark{1,2}}
\affil{Universities Space Research Association, Washington, DC 20024}
\email{olling@usno.navy.mil}

\and

\author{Walter Dehnen\altaffilmark{1}}
\affil{Astrophysikalisches Institut Potdsam, An der Sternwarte 16,
  D-14482 Potdsam, Germany}
\email{wdehnen@aip.de}

\altaffiltext{1}{Equal first author}
\altaffiltext{2}{ 
   Dept. of the Navy, U.S. Naval Observatory, Washington, DC 20392}

\begin{abstract}
  The Oort constants describe the local spatial variations of the
  stellar streaming field. The classic way for their determination
  employs their effect on stellar proper motions. We discuss various
  problems arising in this procedure. A large, hitherto apparently
  overlooked, source of potential systematic error arises from
  longitudinal variations of the mean stellar parallax, caused by
  intrinsic density inhomogeneities and inter-stellar extinction.
  Together with the reflex of the solar motion these variations by
  mode mixing create contributions to the longitudinal proper motions
  $\o{\mu}_{\ell^\star}(\ell)$ that are indistinguishable from the
  Oort constants at $\la20\%$ of their amplitude. Fortunately, we can
  correct for this {\em mode mixing\/} using the latitudinal proper
  motions $\mu_b(\ell)$.
  
  We use about 10$^6$ stars from the ACT/Tycho-2 catalogs brighter
  than $V\approx11$ with median proper motion error of $\approx3\,
  \masyr$. Taking every precaution to avoid or correct for the various
  sources of systematic error, significant deviations from
  expectations based on a smooth axisymmetric equilibrium, in
  particular non-zero $C$ for old red giant stars. We also find
  variations of the Oort constants with the mean color, which
  correlate nicely with the asymmetric drift of the sub-sample
  considered. Also these correlations are different in nature than
  those expected for an axisymmetric Galaxy. 
  
  The most reliable tracers for the ``true'' Oort constants are red
  giants, which are old enough to be in equilibrium and distant enough
  to be unaffected by possible local anomalies. For these stars we
  find, after correction for mode-mixing and the asymmetric drift
  effects, $A\approx16$, $B\approx-17$, $A-B\approx33$, and
  $C\approx-10$\,\kmskpc with internal errors of about 1-2 and
  external error of perhaps the same order. These values are
  consistent with our knowledge of the Milky Way (flat rotation curve
  and $\Omega\equiv A-B\approx28\pm2$).  Based on observations made
  with the ESA Hipparcos astrometry satellite.
\end{abstract}


\section{Introduction}     \label{sec:intro}
\ifpdf
\footnotetext[1]{Equal first author}
\footnotetext[2]{Universities Space Research Association, Washington, DC 20024}
\fi
For over eight decades, the distribution and kinematics of stars in
the solar neighborhood has been studied in order to infer the
structure of the Milky Way galaxy. \citet{kvR20} used star counts to
determine the size and thickness of the Milky Way. Furthermore, by
assuming hydrostatic equilibrium perpendicular to the Galactic plane,
the radial velocities and proper motions of nearby stars allowed
\citet{kap22} to make the first reasonable estimate of the mass of the
Milky Way.  However, \citet{oort27a} pointed out that the mass of
Kapteyn's Galaxy is not large enough to keep the globular clusters and
RR Lyrae stars bound to the Galaxy. \citet{lb27} proposed that the
sub-systems of high-velocity stars and globular clusters as well as
that of nearby low-velocity stars have the same axis of symmetry and a
common center -- that of the globular clusters \citep{hS18}. Lindblad
further hypothesized that each of these sub-systems is in dynamical
equilibrium, and that the sub-systems with the largest amount of
rotation will have the flattest space distribution and the smallest
peculiar velocities \citep{jhJ22}, in agreement with the then
available data.

The motions of the stars in the solar neighborhood can be interpreted
as a streaming (average) velocity plus random motions. In disk
galaxies, the first dominates over the second: such stellar systems
are said to be dynamically cold.  In the solar neighborhood, for
instance, the velocity dispersion in the plane, i.e.\ the amplitude of
random motions, is 45\,\kms\ for the old stellar disk and 18\,\kms\ for
early-type stars, while the streaming motion is of the order of
200\,\kms. In the cold limit of vanishing random motions, the streaming
is along the closed orbits supported by the gravitational potential.
Thus, knowledge about the potential of the Milky Way, and hence its
mass distribution, can be gained from studies of the stellar streaming
velocities.

Oort laid the theoretical basis of this method with his pioneering
paper \citeyearpar{oort27a}. To begin with, let us follow Oort and
consider the cold limit in which all stars move on closed orbits. Oort
himself actually considered the Milky way to be axisymmetric, but his
analysis is easily generalized to non-axisymmetry
\citep{Og32,M35,cha42}. At each position $\B{x}$ in the Galaxy, there
is a unique streaming velocity $\B{v}$ (here, we ignore the
possibility of orbit crossing).  The difference $\delta\B{v}$ between
the velocity at some point in the Galaxy and that at the Sun may be
expanded in a Taylor series (with local Cartesian coordinates:
$\hat{\B{e}}_x$ and $\hat{\B{e}}_y$ pointing in directions
$\ell=0\arcdeg$ and $\ell=90\arcdeg$)
\begin{equation} \label{eq:taylor}
  \delta\B{v} = \mathbf{H} \cdot \B{x} + {\cal O}(\B{x}^2)
\end{equation}
with 
\begin{equation} \label{eq:oort-def-b}
  \mathbf{H} = \left(
    \begin{array}{c@{\;\;}c}
      \partial v_x/\partial x & \partial v_y/\partial x \\ 
      \partial v_x/\partial y & \partial v_y/\partial y
    \end{array} 
  \right)_{\B{\scriptstyle x}=0} \equiv \left(
    \begin{array}{c@{\;\;}c}
      K+C & A-B \\
      A+B & K-C
    \end{array}
  \right).
\end{equation}
The parameters $A$, $B$, $C$, and $K$ are the Oort constants, they
measure the local divergence ($K$), vorticity ($B$), azimuthal ($A$)
and radial ($C$) shear of the velocity field generated by closed
orbits. A star at Galactic longitude $\ell$ and distance $d$ from an
observer and moving with velocity $\B{v}$ relative to the latter, is
observed to have radial and tangential velocity
\begin{mathletters} \label{eq:vel}
  \begin{eqnarray}
    v_d     &=& \phm{+} v_x \cos\ell + v_y \sin\ell \\
    v_\ell  &=& -       v_x \sin\ell + v_y \cos\ell.
  \end{eqnarray}
\end{mathletters}
Inserting \eqn{taylor} with $\B{x}=d(\cos\ell,\sin\ell),$ whereby
assuming the observer participates in the streaming field, one finds
\begin{mathletters} \label{eq:oort}
  \begin{eqnarray} 
    v_d   \,d^{-1} &=& K+A\sin2\ell+C\cos2\ell+\mathcal{O}([d/R_0]^2) \\[0.5ex]
    v_\ell\,d^{-1} &=& B-C\sin2\ell+A\cos2\ell+\mathcal{O}([d/R_0]^2).
  \end{eqnarray}
\end{mathletters}
It is worth emphasizing that terms quadratic in $d/R_0$ contribute to
the $m=\pm1,3$ harmonics $\mathrm{e}^{im\ell}$ in \eqn{oort} and thus
do not interfere with the Oort constants -- however, terms of higher
odd orders do.  Similarly, the deviations of the solar velocity from
the local streaming velocity leads to $m=\pm1$ harmonics, see below.

The Oort constants may also be expressed in terms of cylindrical
coordinates $(R,\varphi)$ with the Sun at $(R_0,0)$ (cf.\ 
Chandrasekhar 1942)
\begin{mathletters} \label{eq:oort-def-a}
  \begin{eqnarray} 
    2A &=& \phm{-} \frac{v_\varphi}{R} - \frac{\partial v_\varphi}{\partial R}
             - \frac{1}{R} \frac{\partial v_R}{\partial\varphi},\\[0.5ex]
    2B &=&   -  \frac{v_\varphi}{R} - \frac{\partial v_\varphi}{\partial R}
             +  \frac{1}{R} \frac{\partial v_R}{\partial\varphi},\\[0.5ex]
    2C &=&   -  \frac{v_R}{R} + \frac{\partial v_R}{\partial R}
             -  \frac{1}{R} \frac{\partial v_\varphi}{\partial\varphi},\\[0.5ex]
    2K &=& \phm{-} \frac{v_R}{R} + \frac{\partial v_R}{\partial R}
             + \frac{1}{R} \frac{\partial v_\varphi}{\partial\varphi}
  \end{eqnarray}
\end{mathletters}
evaluated at the solar position (we use the convention that $\varphi$
increases clockwise, i.e.\ in the direction of Galactic rotation). In
the axisymmetric limit, $C= K= 0$, and
\begin{mathletters} \label{eq:oort-sym}
  \begin{eqnarray}
    A_{\rm sym} &=& \frac{1}{2}\left(
      \phm{-}\frac{v_\varphi}{R}
      -\frac{\partial v_\varphi}{\partial R}\right)_{R=R_0},
    \\[0.5ex]
    B_{\rm sym} &=& \frac{1}{2}\left(
      -\frac{v_\varphi}{R}
      -\frac{\partial v_\varphi}{\partial R} \right)_{R=R_0},
  \end{eqnarray}
\end{mathletters}
equivalent to the equations actually given by Oort. In an axisymmetric
Galaxy, the circular (closed) orbits have velocity $v_\varphi^2=
R\;\partial\Phi/\partial R$, and measurements of the Oort constants
provide a direct constraint on the Galactic potential $\Phi$. For
instance, a harmonic potential results in solid-body rotation, $A=0$,
and $B$ equal to the (constant) rotation frequency, a flat rotation
curve gives $A={-}B$, and the case of all the mass concentrated in the
inner Galaxy yields $A={-}3B$. From the then available radial
velocities and proper motions, \citet{oort27b} found
$A\approx19\,\kmskpc$ and $B\approx-24\,\kmskpc$ (with large
uncertainties, though).  This was clear evidence for a rotation of the
Milky Way (not well established at that time), and ruled out
Lindblad's suggestion that the Milky Way rotates like a solid body.
Note, however, that zero $C$ and $K$ do not necessarily imply
axisymmetry of the Galactic potential; alternatively, the Sun might be
located near a principal axis of an elliptic potential \citep{kt94}.

Given their pre-eminent importance, it should not come as a surprise
that the observational determination of the Oort constants has been
high on the astronomers' priority list.  However, measuring these
streaming velocities directly is no simple task, mainly because of the
lack of an appropriate reference system which is unaffected by
systematic motions and not involved in the Galactic streaming.  For
example, stellar positions and proper motions in the fundamental
stellar catalogs (e.g.\ AGK3, FK4, FK5) are based on transit time
measurements.  The so-determined proper motions are absolute, but with
respect to a reference frame tied to the Earth and the orbits of Solar
system objects, and are thus useful to determine Earth's precession
rate and the motion of the equinox \citep[e.g.,][]{F71,C72,MS93,LK95}.
In order determine the Galactic rotation rate, it is more useful to
determine proper motions relative to an extra-galactic inertial
reference frame.  In more modern proper motion programs, distant
galaxies have been used for this purpose \citep[e.g., NPM, SPM, KSZ,
Bonn, Potsdam, see][for a summary]{HippFrame97}.  However, until
recently their usefulness has been limited by the so-called
``magnitude equation'' problem\footnote{Essentially, accurate position
  determination is hampered by the fact that stellar images are
  non-spherical as a result of telescope tracking errors and the
  non-linearity of the photographic plates (van Altena, private
  communications)}.  Comparison with the Hipparcos data showed that
several of these extra-galactic reference frames were not exactly
inertial.  Some significant residual spins (0.25 - 1.25 mas yr$^{-1}$)
of the pre-Hipparcos coordinate systems were found \citep[see][for a
review]{HippFrame97}, which can lead to systematic errors in the Oort
constant $B$ of 1.2-6\,\kmskpc.
 
So far, we have only considered the cold limit and ignored the effect
of random motions.  However, we will see that this is not suitable for
most stars in the solar neighborhood.  That is, the mean velocity
field of a group of stars differs systematically and significantly in
its divergence, vorticity and shear from the (hypothetical) velocity
field induced by closed orbits.  Moreover, higher-order terms in the
Taylor expansion and other effects imply that the Fourier coefficients
in \eqn{oort} are not identical to the divergence, vorticity, and
shear of the mean velocity field, let alone the closed-orbit velocity
field.  These problems (discussed in some detail in \Sec{prac} with
particular emphasize on proper-motion data) introduce significant
systematic errors in the values derived for the Oort constants.  In
most previous studies, many of these problems have been ignored, a
fact that may well explain the diversity among the values derived for
the Oort constants from different data and by different methods
\citep[see][for reviews]{klb86,om98}. The main objective of the
present paper is, after having recognized these systematic errors, to
avoid them as far as possible both by a careful analysis of the data
and a careful interpretation of the results.  \Sec{tech} details our
analysis technique. In \Sec{analysis}, we motivate our selection of
the ACT catalog and analyze its proper motion data.  The results are
discussed and interpreted in \Sec{disc}.  Finally, \Sec{sum} sums up
and concludes.

For the convenience of authors and readers, the units used throughout
this paper will be consistent and will not always be explicitly given.
Distances are measured in \kpc, velocities in \kms, and frequencies,
such as proper motions and the Oort constants, in \kmskpc. Stellar
parallaxes, denoted by the symbol $\pi$, are considered to be inverse
distances and, consequently, have dimension \kpc$^{-1}$, which
corresponds to measuring the actual parallax angle in
milli-arcseconds. Luminosities and colors are given in magnitudes as
usual.
\section{The Oort Constants in Practice}\label{sec:prac}
\subsection{Deviations from the Cold Limit}\label{sec:adrift}
As already pointed out by \citet{oort28} and later discussed by
\citet{kt94}, only in the cold limit of vanishing random motions can
we interpret the mean streaming velocity $\o{\B{v}}$ as the velocity
$\B{v}$ of closed orbits supported by the Galactic potential. In
general, there is a systematic difference, and we may write
\begin{equation} \label{eq:asym-drift}
  \o{\B{v}} = \B{v} - \B{v}_a
\end{equation}
with the asymmetric drift velocity $\B{v}_a$. The asymmetric drift
expresses the lag of the mean velocity with respect to the local
closed orbit. To explicitly make this distinction, let us write in
analogy to \eq{asym-drift}
\begin{mathletters}
  \begin{eqnarray}
    \o{A}&=&A-A_a, \\
    \o{B}&=&B-B_a, \\
    \o{C}&=&C-C_a, \\
    \o{K}&=&K-K_a.
  \end{eqnarray}
\end{mathletters}
Here, $\o{A}$, $\o{B}$, $\o{C}$, and $\o{K}$ represent the mean
velocity field (its divergence, vorticity and shear) of a group of
stars and might be evaluated from \eqn{oort-def-b} or \eq{oort-def-a}
by replacing $\B{v}$ with $\o{\B{v}}$, while $A_a$, $B_a$ etc.\ follow
upon replacing $\B{v}$ with $\B{v}_a$.

For an axisymmetric Galaxy, there is only an azimuthal component
$v_{a\varphi}$, which, for random motions much smaller than the
rotation velocity, can be approximated by Str\"omberg's relation
\citep[see][eq.~4.34, for a derivation from the Jeans equations]{BT}
\begin{equation} \label{eq:stroemberg}
  2 v_\varphi v_{a\varphi} \simeq \sigma_R^2 \left[
    \frac{\sigma_\varphi^2}{\sigma_R^2} -1 - 
    \frac{\partial\ln(\rho\sigma_R^2)}{\partial\ln R} - \frac{R}{
      \sigma_R^2} \frac{\partial \sigma_{Rz}^2}{\partial z} \right].
\end{equation}
Here, $\B{\sigma}^2$ is the (square of the) velocity dispersion tensor
and $\rho$ the stellar density.  Thus, the asymmetric drift is a
function of the radial velocity dispersion but also depends on the
circular velocity, the axis-ratio of the velocity dispersion
ellipsoid, as well as on gradients in the dispersion and stellar
density.  We might use this relation to estimate the expected effect
on Oort's $A$ and $B$. We refer the reader to \citet{LF89} for a
determination of the radial dependence of the radial and tangential
velocity dispersions.  We proceed by neglecting the radial variation
of the last term in \eq{stroemberg} and by assuming that $\rho$ and
$\B{\sigma}^2$ vary exponentially with scale lengths $R_d$ and
$R_{\sigma^2}$.  We then find:
\begin{equation} \label{eq:oort-ad}
  \frac{\partial \ln v_{a\varphi}}{\partial R} \simeq
  - \frac{1}{ R_{\sigma^2}}
  + \frac{k}{2v_\varphi}\left(\frac{1}{ R_d}{+}  
    \frac{1}{ R_{\sigma^2}}\right)
  - \frac{\partial\ln v_\varphi}{\partial R},
\end{equation}
where we have used $k\,v_{a\varphi}\simeq \sigma_R^2$ with
$k=(80\pm5)\,\kms$ (Dehnen \& Binney 1998, hereafter DB98).

For $R_{\sigma^2}=0.45 R_0$ \citep{LF89}, one finds $A_a\approx0.14
v_{a\varphi}$ and $B_a\approx0.01v_{a\varphi}$, almost independent of
$R_d$, $R_0$, and $v_\varphi$.  That is, $B$ is hardly affected, while
$A_a$ might be as large as 3 for red stars (using $v_{a\varphi}$ of 20
according to DB98's findings).

This derivation of $A_a$ and $B_a$ is based on the assumption of
dynamical equilibrium, and hence not appropriate for young stellar
populations, whose lumpy phase-space structure (moving groups and
inhomogeneous spatial distribution) indicates non-equilibrium.
Further, note that the numerical values derived above from
\eqn{oort-ad} are only valid in the Solar neighborhood as $k$ and
$v_{a\varphi}$ were determined locally.

It is important to notice that $\o{A}$, $\o{B}$, $\o{C}$, and $\o{K}$
are functions of Galactic height $z$ (already because the asymmetric
drift depends on $z$). One might think that one could just extend the
Taylor expansion \eq{taylor} into the vertical direction. However,
this does not work, mainly because the scale height of the stellar
disk, and hence any possible kinematic gradients, is much smaller than
reasonable sample depths, i.e.\ high-order terms are significant.
(Also, for symmetry reasons, the first non-trivial term is quadratic
and its distance dependence would not nicely vanish in the resulting
proper motions.)

\subsection{Deviations from Axisymmetry} \label{sec:dev-axi}
For an axisymmetric Galaxy, as originally considered by Oort, the
closed orbits are circular and at each radius there exists exactly one
such orbit.  However, the Galaxy is not axisymmetric, it has a central
bar and its disk appears to have 4, or a 4+2, armed spiral pattern
\citep{V95,AL97}, but the infrared light, originating mainly from old
stars, is dominated by a 2-arm mode \citep{drim00,ds01}. As a
consequence, the closed orbits are no longer circular but elliptical.
The orientation of these elliptic orbits changes when one crosses the
co-rotation or inner and outer Lindblad resonances (CR, ILR, \& OLR)
of the bar or spiral pattern.  Certain places near these resonances
are visited by two or more differently oriented closed orbits. It has
even been proposed that the Sun is precisely at such a position on the
OLR \citep{Kal91}. However, it seems now more likely from detailed
models of the gaseous and stellar motions in the inner parts of the
Milky Way that we are about 1\,\kpc\ or even less outside the OLR
\citep{EG99,Fux99,d99,d00}. The spiral pattern of the Milky Way also
imposes non-axisymmetric perturbations on the velocity field
\citep{LYS69,MPS79}. Analyses of radial velocities and now also of
proper motions of several young tracer populations indicate that the
Sun is located close to the radius of co-rotations of the spiral
density wave \citep[$\Delta R=\pm1$ kpc,
$\Delta\Omega=\pm2\,\kmskpc$;][]{CM73,AL97,MZ99}.

Clearly, the Oort constants, defined in terms of the closed-orbit
velocity field, are ill-defined at these positions, and will behave
discontinuously when crossing one of the three major resonances.  Of
course, in reality one measures the Oort constants from stars that are
not on closed orbits.  In this case, the discontinuities of the Oort
constants will be replaced by a more gradual transition.

\subsection{Measuring the Oort Constants} \label{sec:measure}
The Oort constants are measured by determining the Fourier
coefficients in \eqn{oort}. However, in reality the Galaxy is not flat
but three-dimensional, and we cannot use this equation directly. To
generalize for stars out of the plane, we use
\[ \begin{array}{r@{\,=\,}l}
  \pi    & d^{-1}\,\cos b    \\
  v_r    & v_d\,\cos b       \\
  \o{\mu}_{\ell^\star}   & \pi\,v_\ell       \\
  \mu_b  &-\pi\,v_d\,\sin b,     
\end{array} \]
where $\pi$ denotes the star's parallax, $d$ the distance projected
onto the Galactic plane and $\o{\mu}_{\ell^\star} \equiv \mu_\ell\,
\cos b$ (the quantity actually measured in astrometry). Moreover, the
Sun is not moving with the local streaming velocity but with some
velocity $\o{\B{v}}_\odot \equiv(\o{U},\o{V},\o{W})$ relative to it.
We thus finally have for the observable proper motions, neglecting
higher-order terms in \eqn{taylor},
\ifpreprint
\begin{mathletters} \label{eq:four}
  \begin{eqnarray}
    \label{eq:four-a}
    \o{\mu}_{\ell^\star} &\,=\,& 
    \o{\mu}_U\sin \ell - \o{\mu}_V\cos \ell
    \nonumber\\
    &     &+\cos b\,(\tilde{B}-\tilde{C}\sin 2\ell+\tilde{A}\cos 2\ell)
    \\
    \o{\mu}_b            &\,=\,& 
    \sin b\, (\o{\mu}_U\cos \ell + \o{\mu}_V\sin \ell) -\o{\mu}_W \cos b
    \nonumber\\
    &     &-\sin b\,\cos b\,(\tilde{K}+\tilde{A}\sin 2\ell+\tilde{C}\cos 2\ell),
  \end{eqnarray}
\end{mathletters}
\else
\begin{mathletters} \label{eq:four}
  \begin{eqnarray} 
    \label{eq:four-a}
    \o{\mu}_{\ell^\star} &\,=\,& \o{\mu}_U\sin \ell - \o{\mu}_V\cos \ell
    +\cos b\,(\tilde{B}-\tilde{C}\sin 2\ell+\tilde{A}\cos 2\ell)             \\
    \o{\mu}_b &\,=\,& \sin b\, (\o{\mu}_U\cos \ell + 
    \o{\mu}_V\sin \ell) -\o{\mu}_W \cos b
    -\sin b\,\cos b\,(\tilde{K}+\tilde{A}\sin 2\ell+\tilde{C}\cos 2\ell),
  \end{eqnarray}
\end{mathletters}
\fi
with the reflex of the solar motion
\begin{equation} \label{eq:pm-sun}
  \o{\B{\mu}}_\odot
  \equiv(\o{\mu}_U,\o{\mu}_V,\o{\mu}_W)
  \equiv\pi\,\o{\B{v}}_\odot.
\end{equation}
Note that we have written $\tilde{A}$ instead of $\o{A}$ etc.; the
symbols with a tilde are defined by \eqn{four}, i.e.\ as Fourier
coefficients of the mean proper motion for stars at the same distance,
while those with a bar are the divergence, vorticity, and shear of the
mean velocity field for a group of stars. Strictly speaking, these are
two different sets of quantities, and we can only hope that they are
not too different. Of course, all we can hope to measure are
$\tilde{A}$, $\tilde{B}$, $\tilde{C}$, and $\tilde{K}$, while we want
$\o{A}$, $\o{B}$, $\o{C}$, and $\o{K}$ (actually, $A$, $B$, $C$, and
$K$). We now discuss potential sources of discrepancies between these
two sets of quantities, and further problems in measuring the Fourier
coefficients.

\subsection{Systematics with Sample Depth} \label{sec:sys-depth}
Equations \eq{oort-def-a} are based on the first-order Taylor series
\eq{taylor} and neglect higher-order terms, which might become
important for deep samples. Including the next order adds six new
unknowns, while only the $m=$1, 3 Fourier terms in
$\o{\mu}_{\ell^\star}$ are affected, i.e.\ only two more constraints
are available (the $\cos3\ell$ and $\sin3\ell$ terms).  Thus, the next
higher-order expansion is not generally soluble from proper motion
data alone.

For illustration, we now consider the effect of a purely axisymmetric
model, which adds only one additional unknown per Taylor order
\citep{PMB94,fw97}. To next order in $d/R_0$, we then get at $b=0$:
\ifpreprint
  \begin{eqnarray}
    \o{\mu}_{\ell^\star}  &=& (\o{\mu}_{\ell^\star})_0
    +\frac{d}{8R_0}\left(3R_0
      \frac{\partial^2v_\varphi}{\partial R^2}\Big|_{R_0}
      -6A\right) \cos\ell
    \nonumber \\
    && + \frac{d}{8R_0}\left(R_0
      \frac{\partial^2v_\varphi}{\partial R^2}\Big|_{R_0}
      +6A \right) \cos3\ell, \nonumber
  \end{eqnarray}
\else
  \begin{displaymath}
    \o{\mu}_{\ell^\star}  = (\o{\mu}_{\ell^\star})_0
    + \frac{d}{8R_0}\left(3R_0
      \frac{\partial^2v_\varphi}{\partial R^2}\Big|_{R_0}
      -6A\right)\cos\ell
    + \frac{d}{8R_0}\left(~R_0
      \frac{\partial^2v_\varphi}{\partial R^2}\Big|_{R_0}
      +6A\right) \cos3\ell,
  \end{displaymath}
\fi
where $(\o{\mu}_{\ell^\star})_0$ refers to the expression \eq{four-a}.
From analyses of Cepheid kinematics, the value of $\partial^2
v_\varphi / \partial R^2|_{R_0}$ is estimated to be smaller than
3\,km\,s$^{-1}$\,kpc$^{-2}$ in modulus. With this estimate and
$A\approx15$, one finds that the term ${\cal O}(x)$ contributes up to
$-2.5$ and 1.8 to the $\cos\ell$ and $\cos3\ell$ Fourier terms,
respectively, for $d/R_0=1/8$. Thus, even for a rather smooth
streaming field higher-order terms may become important already at
modest sample depth.

\subsection{Effects of a Non-Smooth Streaming Field} \label{sec:sys-r}
Because of local anomalies in the Galactic force field, e.g.\ caused
by spiral arms and other distortions, the streaming field {\em
  inevitably} exhibits small-scale oscillations on top of an
underlying smooth field. These oscillations give rise to significant
higher-order terms in the Taylor expansion \eq{taylor}. Since in
practice the Oort constants are measured from the kinematics of large
stellar samples with a finite depth, these high-order terms become
important. In the present context, however, we are predominantly
interested not in the small-scale but the large-scale behavior of the
streaming field. When using a deep sample, i.e.\ a big volume, one
might hope that the small-scale wiggles average out and one is left
with the Oort constants due to the smooth equilibrium field.
\begin{figure}[th]
   \ifpreprint
      \centerline{\resizebox{85mm}{!}{\includegraphics{f1}}}
   \else
      \centerline{\resizebox{70mm}{!}{\plotone{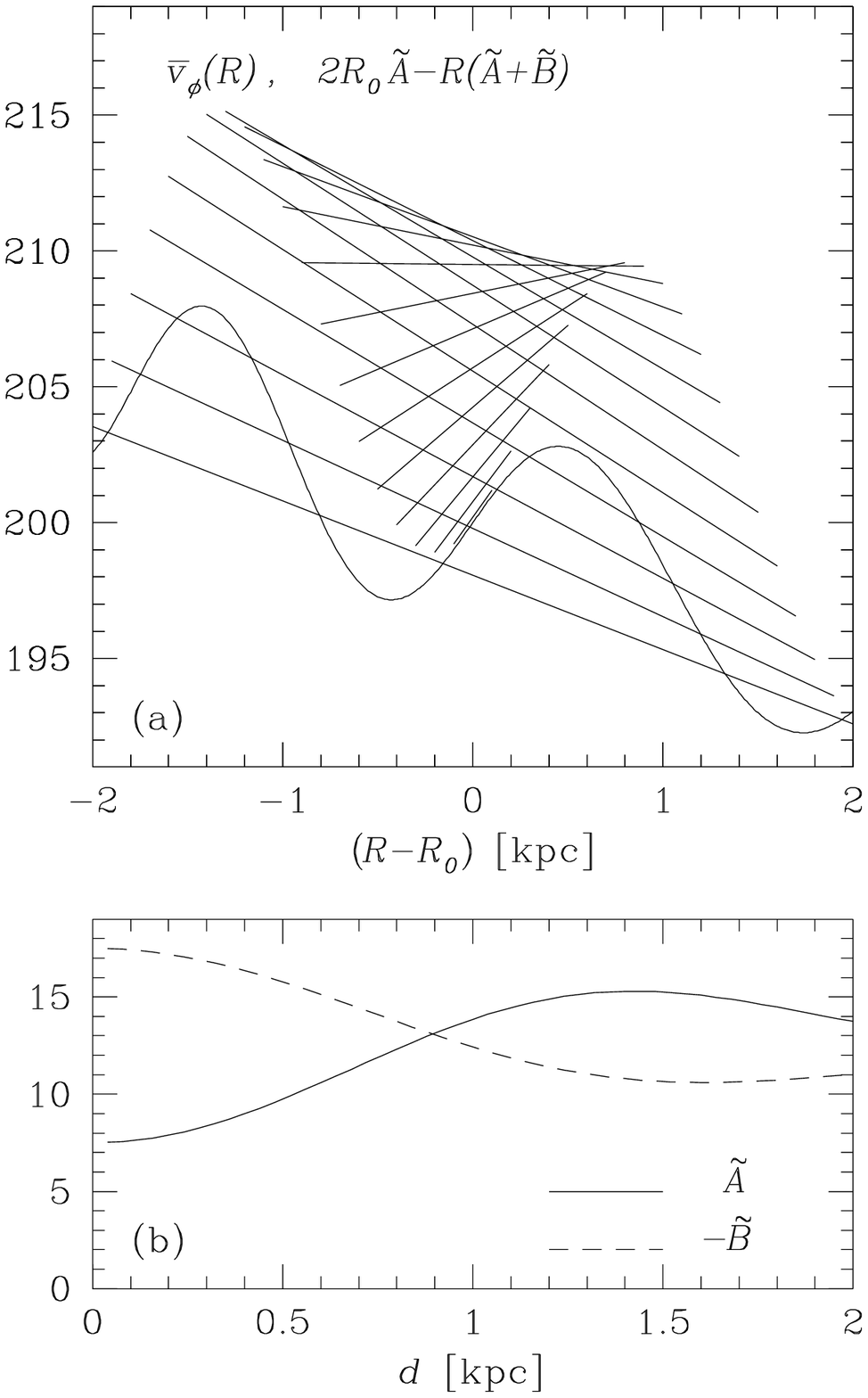}}}
   \fi
 \input{ms_fc01}
\end{figure}
We have simulated this effect in \Fig{wiggle}, where an axisymmetric
streaming velocity (black line in upper panel) with wiggles of
wave-length $\lambda\approx2\,\kpc$ and amplitude of only {\em two}
percent of an otherwise smooth (power-law) rotation velocity is
assumed.  From this model, and the {\em exact} proper motion equation
\begin{equation} \label{eq:mul_cold_axi}
  \mu_\ell(\ell,d) = \frac{ \o{v}_\varphi(R) [R_0\cos \ell-d] / R
    - \o{v}_\varphi(R_0)\cos \ell }{d},
\end{equation}
we measured the Fourier coefficients $\tilde{A}$ and $\tilde{B}$ as
function of distance $d$ (\fig{wiggle}b).  Assuming that the
$\tilde{A}$ and $\tilde{B}$ obtained in this way measure the local
streaming field, one may approximate that locally
\begin{displaymath}
  \o{v}_\varphi \simeq 2R_0\tilde{A} - R(\tilde{A}+\tilde{B}).
\end{displaymath}
The straight lines in \fig{wiggle}a, reaching from $R_0-d$ to $R_0+d$,
represent these approximations.  One finds that for nearby stars
($d\ll\lambda$) the measured Oort constants $\tilde{A}$ and
$\tilde{B}$ accurately represent the (wiggly) local streaming field.
For distances $d$ similar to or larger than $\lambda$, $\tilde{A}$ and
$\tilde{B}$ represent the underlying smooth part.  For intermediate
distances, there is a transition with a range in
$R_0(\tilde{A}-\tilde{B})$ [the estimate for $\o{v}_\varphi(R_0)$]
comparable to twice the amplitude of the wiggles.  The estimated
gradient of $\o{v}_\varphi$ (the slope of the straight [red] lines) is
a strong function of $d$, giving sensible results only for
$d>\lambda$.

The importance of these effects is likely to be strongest for young
stars and weakest for old stars, because their larger velocity
dispersion makes old stellar populations less susceptible to
small-scale features in the force field.  From the above it is clear
that even the smallest degree of non-smoothness (2\%) in the Galactic
force field can have a significant effect on the apparent values of
the Oort constants (30\%).  In fact, one can show that the degree to
which a stellar population with radial dispersion $\sigma_R$ responds
to small-scale perturbations is inversely proportional to $\sigma^2_R$
(\cite{M74}).  Thus, while young stars are ideally suited to trace
Galactic fine-structure, old stars with their larger dispersion are
much less influenced by any wiggles in the force field and are thus
more suitable to study the large-scale potential of the Milky Way.  A
quantitative estimate for the critical wave-length to which a group of
stars with dispersion $\sigma_R$ is just sensitive, may be given by
the average epicycle diameter $\sqrt{8}\sigma_R/\kappa$.  With a
epicycle frequency of $\kappa\approx36\,\kmskpc$ and $\sigma_R$ of
$\approx15\,\kms$ and $\approx38\,\kms$ for blue and red stars (DB98), we
find a limiting wave-length of $\approx1\,\kpc$ and $\approx3\,\kpc$,
respectively.  From the structure in the interstellar medium (ISM), we
expect wiggles on the scale of about 2\,\kpc\ \citep{om98}, which can
result in 30\% ``errors'' in the Oort constants.

Groups of stars with a radial velocity dispersion larger than about
25\,\kms\ have an epicyclic diameter larger than the wavelength of the
ISM-induced wiggle of the rotation curve.  Main-sequence stars bluer
than $\bv=0.42$ have dispersions smaller than 25\,\kms\ (DB98) and
main-sequence lifetimes less than about 2~Gyr.  Thus, the kinematics
of stars younger than 2~Gyr may be significantly influenced by
low-amplitude, small-scale perturbations in the Galactic potential.
Non-axisymmetric perturbations --which were not considered by OM98--
could lead to even larger differences between the apparent Oort
constants and the ``true'' Oort constants.

In fact, the situation depicted in \Fig{wiggle} is likely to be a
simplification.  In the calculations leading to \fig{wiggle} we
assumed that the Sun is located half-way between the extrema of the
rotation curve wiggle.  That is to say, we assumed that the Sun is
located in a special place with respect to the inevitable small-scale
oscillations of the Galactic force field. Olling \& Merrifield's
(1998) work suggests that, indeed, the Sun is not located in a ``sweet
spot'' of the wiggly $\o{v}_{\varphi}(R)$ curve \citep[see
also][]{AOLM96}. In that case, the widely employed technique of
expanding the rotation curve and the velocity field to low-order 
(\ref{eq:taylor}) is not adequate.

And finally, since there is evidence that the Milky Way exhibits
significant azimuthal asymmetry (\Sec{dev-axi}), the arguments
presented above are likely to be over-simplifications.  Nevertheless,
the reasoning above gives us an indication as to the level of possible
systematic errors in our analyses (see also \Sec{Systematic_Errors}).

\subsection{Correlations of Parallax with Kinematics} \label{sec:sys-corr}
In reality, of course, the stars are not just at one distance, but
distributed over the sampling volume.  That is, for each $\ell$, one
averages along the line of sight.  In order to arrive at an equivalent
to \eqn{four} for a sample of stars are various $\pi$, one must assume
that $\o{\pi\,U}=\o{\pi}\,\o{U}$ and analogously for $\o{\pi\,V}$ and
$\o{\pi\,W}$, i.e.\ that parallaxes and velocities are uncorrelated.
However, for stars associated with spiral arms one expects such
correlations.  A systematic error is introduced when using this
assumption for stellar samples for which it is not justified.

\subsection{Mode Mixing} \label{sec:mix}
When measuring the Oort constants from proper motions, one often uses
a magnitude-limited sample with little information about the stellar
parallaxes.  One then has to replace $\pi$ in \eq{pm-sun} by the
line-of-sight averaged parallax $\o{\pi}$ (or, more precisely,
$\pi\o{\B{v}}_\odot$ by $\o{\pi\B{v}}_\odot$). Due to non-uniformity
both of the stellar space-density and of the extinction, $\o{\pi}$
will {\em inevitably\/} depend on longitude. We might expand it into a
Fourier series in $\ell$
\begin{equation} \label{eq:mean-pi}
  \o{\pi} \equiv \o{\pi}_0 \Big( 1 + 2 
  \sum_m \cp{m}\cos m\ell + \sp{m}\sin m\ell \Big),
\end{equation}
where the pre-factor of two in front of the Fourier sum has been
introduced for later convenience. Inserting \eq{mean-pi} and
\eq{pm-sun} into \eqn{four} gives
\begin{mathletters} \label{eq:mix-mu} 
  \begin{equation} \label{eq:mix-mul}
    \begin{array}{rclcclclclcl}
      \o{\mu}_{\ell^\star} &=&
                  &(& &\tilde{B}\cos b&+&\sp1\o{\mu}_{U0}&-&\cp1\o{\mu}_{V0})\\
      &+&\cos\ell &(&-&\o{\mu}_{V0}   &+&\sp2\o{\mu}_{U0}&-&\cp2\o{\mu}_{V0})\\
      &+&\sin\ell &(& &\o{\mu}_{U0}   &-&\cp2\o{\mu}_{U0}&-&\sp2\o{\mu}_{V0})\\
      &+&\cos2\ell&(& &\tilde{A}\cos b&-&\sp-\o{\mu}_{U0}&-&\cp+\o{\mu}_{V0})\\
      &+&\sin2\ell&(&-&\tilde{C}\cos b&+&\cp-\o{\mu}_{U0}&-&\sp+\o{\mu}_{V0})\\
      &+&\dots, & & & & & & &
    \end{array}
  \end{equation} 
\ifpreprint
  \begin{equation} \label{eq:mix-mub}
    \begin{array}{rclccl}
     \o{\mu}_b &=&
     &(&-&\cos b\,[\o{\mu}_{W0}+\sin b\tilde{K}]                         \\
     & &         & &+&\sin b\,[\cp1\o{\mu}_{U0}+\sp1\o{\mu}_{V0}])       
     \\[0.4ex]
     &+&\cos\ell &(&-&2\cp1\cos b\,\o{\mu}_{W0}                          \\
     & &         & &+&\sin b[\o{\mu}_{U0}+\cp2\o{\mu}_{U0}+\sp2\o{\mu}_{V0}])
     \\[0.4ex]
     &+&\sin\ell &(&-&2\sp1\cos b\,\o{\mu}_{W0}                          \\
     & &         & &+&\sin b\,[\o{\mu}_{V0}+\sp2\o{\mu}_{U0}-\cp2\o{\mu}_{V0}])
     \\[0.4ex]
     &+&\cos2\ell&(&-&2\cp2\cos b\,\o{\mu}_{W0}                          \\
     & &         & &+&\sin b\,[-\cos b\,\tilde{C}+\cp+\o{\mu}_{U0}
                                                 -\sp-\o{\mu}_{V0}])
     \!\!\!\!\!\!\!\!\!\! \\[0.4ex]
     &+&\sin2\ell&(&-&2\sp2\cos b\,\o{\mu}_{W0}                          \\   
     & &         & &+&\sin b\,[-\cos b\,\tilde{A}+\sp+\o{\mu}_{U0}
                                                 +\cp-\o{\mu}_{V0}])
     \!\!\!\!\!\!\!\!\!\! \\[0.4ex]
     &+&\dots,   & & &
    \end{array}
  \end{equation} 
\else
  \begin{equation} \label{eq:mix-mub}
    \begin{array}{rclcclcl}
     \o{\mu}_b &=&
     &(&-&\cos b\,[\o{\mu}_{W0}+\sin b\tilde{K}]
                   &+&\sin b\,[\cp1\o{\mu}_{U0}+\sp1\o{\mu}_{V0}])
                   \\
     &+&\cos\ell &(&-&2\cp1\cos b\,\o{\mu}_{W0}
                   &+&\sin b[\o{\mu}_{U0}+\cp2\o{\mu}_{U0}+\sp2\o{\mu}_{V0}])
                   \\
     &+&\sin\ell &(&-&2\sp1\cos b\,\o{\mu}_{W0}
                   &+&\sin b\,[\o{\mu}_{V0}+\sp2\o{\mu}_{U0}-\cp2\o{\mu}_{V0}])
                   \\
     &+&\cos2\ell&(&-&2\cp2\cos b\,\o{\mu}_{W0}
                   &+&\sin b\,[-\cos b\,\tilde{C}+\cp+\o{\mu}_{U0}
                                                 -\sp-\o{\mu}_{V0}])
                   \\
     &+&\sin2\ell&(&-&2\sp2\cos b\,\o{\mu}_{W0}
                   &+&\sin b\,[-\cos b\,\tilde{A}+\sp+\o{\mu}_{U0}
                                                 +\cp-\o{\mu}_{V0}])
                   \\
     &+&\dots,   & & & & &
    \end{array}
  \end{equation} 
\fi
\end{mathletters}
where $\cp\pm\equiv\cp1\pm\cp3$, $\sp\pm\equiv\sp1\pm\sp3$,
$s_{\pi,2\pm4}\equiv s_{\pi2} \pm s_{\pi4}$ and
\begin{equation} \label{eq:pm-sun-zero}
  \o{\B{\mu}}_{\odot0}\equiv(\o{\mu}_{U0},\o{\mu}_{V0},\o{\mu}_{W0})
  \equiv\o{\pi}_0 \B{v}_\odot.
\end{equation}
Thus, the $m=0$ and $m=2$ Fourier coefficients of
$\o{\mu}_{\ell^\star}$ measurable for a stellar sample are {\em not
  identical\/} to $\tilde{A}$, $\tilde{B}$, and $\tilde{C}$, rather
mode mixing leads to additional contributions from the solar reflex
motion.  Likewise, and contrary to the classical no-mode-mixing case,
the Solar reflex motion terms ($m=1$) have contributions from
$\o{\mu}_{V0}$\ {\em and} $\o{\mu}_{U0}$.

\subsubsection{The Size of the Effect} \label{sec:mix.size}
It is instructive to estimate the possible size of the effect under
realistic conditions like those we will encounter in our application
to proper-motion catalogs below. Let us assume that we have a stellar
sample at low latitudes with mean parallax $\o{\pi}$ of 2 mas,
corresponding to a typical depth of about 500 pc. Then $m=1,3$
amplitudes of the mean parallax of only 10\% causes a contribution of
about 1-3\,\kmskpc\ to the observable $m=0,2$ harmonic of
$\o{\mu}_{\ell^\star}$, i.e.\ the Oort constants (with
$\o{U}\approx10\,\kms$, $\o{V}\approx20\,\kms$). This is larger than the
uncertainty of the raw Fourier coefficients. Thus, mode mixing
dominates the error of the Oort constants when determined from proper
motions surveys, unless the $m=1,3$ amplitudes of $\o{\pi}$ are much
smaller than 10\%\footnote{Of course, if the sample is deeper, the
  effect is alleviated. However, then other unwanted effects appear,
  see \Sec{sys-r}.}.  Most amazingly, the corresponding literature is
absolutely void of any remarks on this nasty effect.
\begin{figure}[th]
   \ifpreprint
      \centerline{\resizebox{90mm}{!}{\includegraphics{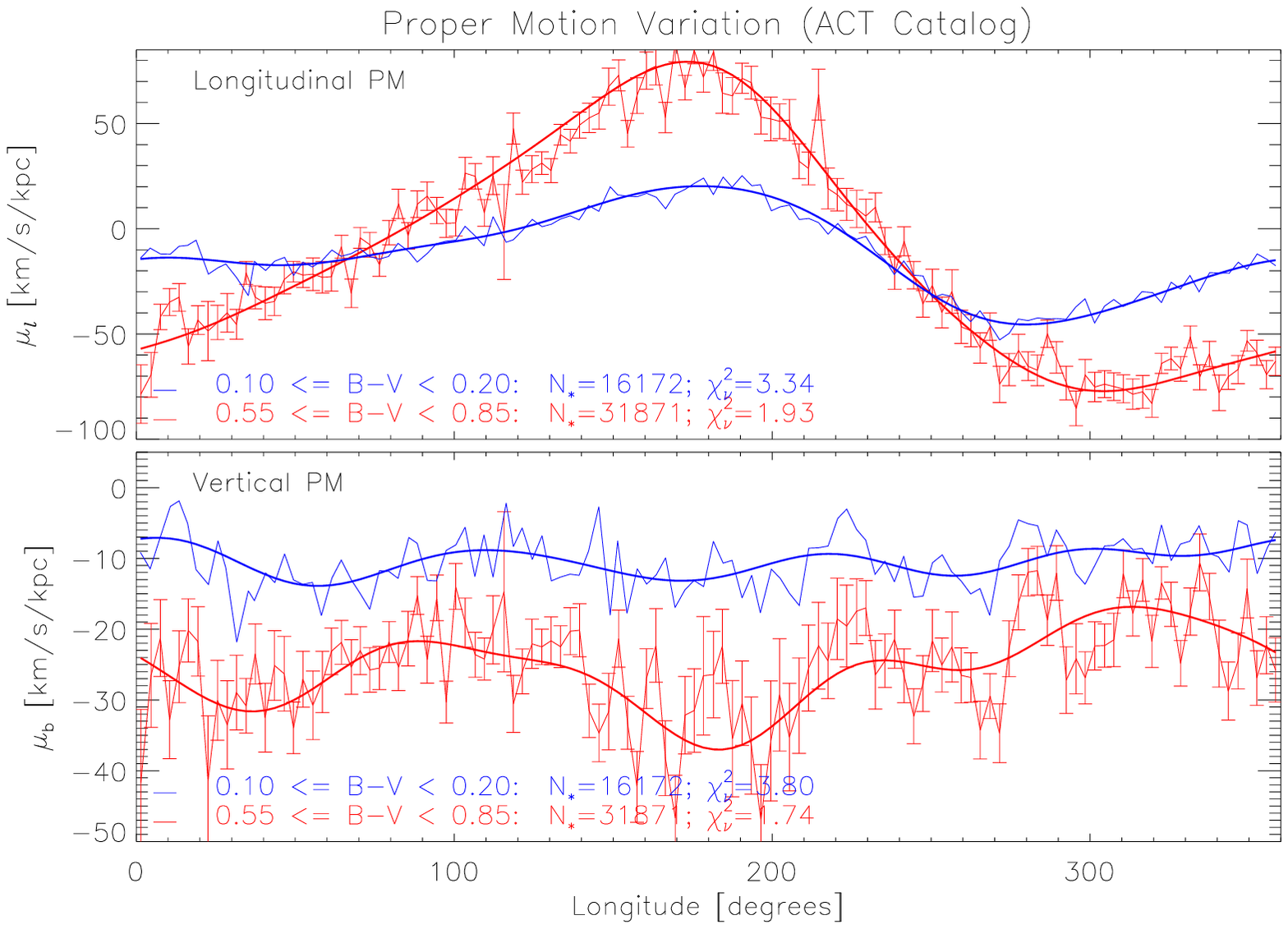}}}
   \else
      \centerline{\resizebox{75mm}{!}{\plotone{f2}}}
   \fi
   \input{ms_fc02}
  \end{figure}
One may show that the longitudinal dependence of the mean parallax
induced just by a smooth exponential stellar disk is
small\footnote{For a (volume-complete) sample with depth $d_{\rm max}$
  in an exponential disk with scale length $R_d$, $s_{\pi m}\equiv0$
  and, to first order in $d_{\rm max}/R_0$,
  \begin{equation} \label{eq:mean-pi-exp}
    c_{\pi1}\o{\pi}_0 = - \frac{\cos b}{16R_d},
    \qquad c_{\pi3}\o{\pi}_0 = 0.
  \end{equation}
  This leads to mode-mixing contributions of 0.25 to 0.6 for early to
  late stellar types, respectively, independent of $\o{\pi}_0$ (at
  $b=0$ and with $R_d=2.5\,\kpc$, $\o{U}\approx10$, and $\o{V}$ between
  10 and 25, DB98).}.  However, as we will see below, the stars are
very non-uniformly distributed in $\ell$. This is a clear indication
of inhomogeneity in the spatial distribution of the sample stars.
This inhomogeneity is caused both by an intrinsic clumpiness of the
stellar density and by extinction blocking the view through the Galaxy
in a highly inhomogeneous and anisotropic way.  Thus, we expect
$\o{\pi}$ to be non-uniform as well, causing considerable mode mixing.


In \Fig{muL_muB} we present graphically the longitudinal variation of
$\o{\mu}_{\ell^\star}$ and $\mu_b$ for two groups of stars with
different colors. The ``jagged'' lines represents the data, the smooth
lines the Fourier fits. We clearly see that the vertical proper motion
exhibits azimuthal variation of about $\pm$ 35\% (bottom panel),
indicating a changing average parallax with longitude. The reduced
$\chi^2$ values listed are computed with respect to a 5$^{th}$ order
Fourier-fit model and indicate that the fits are ``reasonable.''
However, because the distribution of the proper motions in each
longitude bin of 3$^o$ is not quite normal, the $\chi_\nu^2$ values
are only indicative of the goodness of fit.  Assuming a simpler model
with constant $\mu_b(\ell)$ lead to 60\% larger $\chi_\nu^2$ values,
proving that the observed variation is real. The mean levels of the
vertical proper motion indicate that the red stars are about three
times closer than the blue stars.

\subsubsection{A Cure from the Effect?} \label{sec:mix.cure}
In order to correct for this effect, one needs some unbiased estimate
of the relative Fourier coefficients $(c_\pi,s_\pi)_m$ for the mean
stellar parallax, defined in \eqn{mean-pi}.

Fortunately, for $b=0$ \eqn{mix-mub} simplifies considerably to
\begin{equation} \label{eq:mix-mub-low}
  \o{\mu}_b|_{b=0} = - \o{\mu}_{W0} \Big( 1 + 2 
  \sum_m \cp{m} \cos m\ell + \sp{m} \sin m\ell \Big).
\end{equation}
Thus, for low latitudes we can use the vertical stellar proper motions
to measure, apart from $\o{\mu}_{W0}$, the relative Fourier
coefficients $(c_\pi, s_\pi)_m$ of $\o{\pi}$ and use them to correct
the Fourier coefficients obtained from $\o{\mu}_{\ell^\star}$ (see
\Sec{tech-mix} for details of our measurement technique).  However,
this works only, if the vertical solar reflex motion $W_0$, or
equivalently, the mean vertical motion of the stars with respect to
the Sun is constant with $\ell$. A stellar warp, for example, would
invalidate this assumption. Fortunately, it seems from Dehnen's (1998)
analysis of the local stellar velocity distribution that the stellar
warp only weakly affects our analysis: only stars at $R- R_0\ga1\,\kpc$
or with high $V$-velocity are affected.

If this assumption of constant $W_0$ is doubtful, or if high-latitude
stars are concerned, the only possible cure of the problem is an
unbiased, but not necessarily very accurate, distance measure, which
can be used to independently estimate the Fourier coefficients of
$\o{\pi}$. Using the photometric parallaxes for this purpose does not
work -- we tried it -- presumably since extinction affects the
distributions of true and photometric parallax in fundamentally
different ways, but perhaps also because of systematic errors in the
photometric parallax (cf.\ Malmquist bias).

\subsubsection{Mode Mixing and Distance Effects} \label{sec:mix.dist}
In \Sec{sys-depth} we have seen that for a conventional
parameterizations of the average Galactic streaming field, distance
effects result in $\cos{\ell}$ and $\cos{3\ell}$ amplitudes of order
of several \kmskpc.  Such contributions are comparable to those
expected from mode-mixing effects (\Sec{mix.size}).  Thus, for samples
that are sufficiently deep for these distance effects to be important
one cannot disentangle the two contributions, and a correction of the
mode mixing from proper motion data alone becomes impossible.  We will
return to this point in the discussion of the results obtained from
proper-motion data.

\subsection{Measurement Accuracy} \label{sec:accur}
\subsubsection{Internal Uncertainties}
It is instructive to estimate the expected uncertainty for
$\tilde{A}$, $\tilde{B}$, and $\tilde{C}$ determined from the proper
motions of a large stellar sample ($\tilde{K}$ can only be sensibly
measured from radial velocities). The observed proper motion for an
individual star can be described by the mean plus a random component
and an error term. The mean as function of $(\ell,b)$ is given by
\eqn{four} (neglecting mode mixing). The random component has two
sources: random stellar motions leading to a random component
proportional to parallax, and scatter in stellar parallaxes resulting
in scatter in the terms arising from the solar reflex motion.

For a stellar sample uniformly distributed inside a radius $d_{\rm
  max}$, with mean motion with respect to the Sun of 20\,\kms, and a
one-dimensional velocity dispersion in the plane of 30\,\kms, the
dispersion of the random proper motion amounts to roughly $50/d_{\rm
  max}\,\kmskpc$. Thus, in order to measure the Oort constants to an
accuracy of $\epsilon$, we need of the order of $N=2500\times d_{\rm
  max}^{-2}\times(\epsilon/\kmskpc)^{-2}$ stars. If one increases the
accuracy by using ever deeper samples, one must be aware that, as
discussed in \Sec{sys-r} above, the measurable Oort constants are
functions of sample depth.  From this estimate, it is clear that the
measurement uncertainties in the observed proper motions are almost
always negligible in the error budget.

\subsubsection{Systematic Errors} \label{sec:Systematic_Errors}
There is, however, another well known source of uncertainty: a
residual systematic rotation of the astrometric reference system used
to measure the individual proper motions.  The rotation frequency
simulates a vorticity of the velocity field, and its $z$ component is
indistinguishable from $\tilde{B}$.  For ground-based proper-motion
surveys, one can only with great difficulty establish an inertial
reference frame, and in many cases some residual rotation remains
\citep[see][for a recent review]{HippFrame97}. For example, the
rotation of the Hipparcos reference frame is estimated to be no larger
than 0.25 mas\,yr$^{-1}=1.18\,\kmskpc$ \citep{HippFrame97}.  This
implies that $B$ cannot be accurately measured, even when using the
best astrometric reference system currently available!
\section{The Measurement Technique} \label{sec:tech}
The classical way used to determine the Oort constants from proper
motions $\B{\mu}\equiv(\o{\mu}_{\ell^\star} , \mu_b)$ is a
least-squares fit of \eqn{four} to the observed proper motions.
\citet{hanson87}, for instance, employed this method to determine just
$A$ and $B$ from the NPM survey.  This straightforward approach,
however, cannot be recommended.  The main problem is that the actual
functional form of the mean proper motion $\o{\B{\mu}}(\ell,b)$ is
possibly not well described by the fitting function (see \Sec{prac}).
As with all parametric methods, such a mismatch inevitably leads to
systematic errors in the values obtained for the parameters.

As already discussed in \Sec{adrift} and \Sec{measure} above, the
measurable Oort constants and the relative mean proper motions
$\o{\mu_U}, \,\o{\mu_V},\,\o{\mu_W}$ are functions of stellar type and
latitude.  These variations introduce an a priori unknown dependence
of $\o{\B{\mu}}$ on latitude $b$.  In his analysis Hanson accounted
for a variation of $\o{\B{\mu}}_\odot$, but did not find a latitude
dependence of $A$ and $B$.  Additional to the expected variations with
latitude, there will be variations of $\o{\B{\mu}}$ with longitude as
discussed in \Sec{sys-depth} and \Sec{sys-r}.  The stars are always
non-uniformly distributed in $\ell$ such that the harmonics
e$^{im\ell}$ are not orthogonal under averaging over $\ell$.  This
implies that, if there is (non-anticipated) power in higher-order
harmonics, fitting only the low-order ($m\le2$) harmonics again
introduces systematic errors in the parameter values obtained.

In order to avoid these problems, we expand $\o{\B{\mu}}$ in a Fourier
series including high orders and restrict ourselves to a small range
of latitudes.  This is an essentially non-parametric approach in that
it allows for a general functional form for $\o{\B{\mu}}(\ell)$ and
avoids variations at high latitudes.  To be more specific we determine
the coefficients $c_{\ell m}$, $s_{\ell m}$, $c_{b m}$, and $s_{b m}$
of
\begin{mathletters} \label{eq:mu-fit}
  \begin{eqnarray} \label{eq:mul-fit}
    \o{\mu}_{\ell^\star}    &=& c_{\ell0} + \sum_{m>0}^{m_{\rm max}} 
    c_{\ell m} \cos m\ell + s_{\ell m} \sin m\ell, \\
    \label{eq:mub-fit}
    \o{\mu}_b    &=& c_{b0} + \sum_{m>0}^{m_{\rm max}} 
    c_{b m} \cos m\ell + s_{b m} \sin m\ell
  \end{eqnarray}
\end{mathletters}
by a least-squares fit to the observed $\o{\mu}_{\ell^\star}$ and
$\mu_b$ (see also \Sec{tech-num-raw} below).

\subsection{Weighting and Exclusion of Outliers} \label{sec:weight}
In order to minimize the errors, one might weight individual stars
with the inverse variance of their proper motion. This is essentially
given by the stars parallax times the velocity dispersion, see
\Sec{accur} above. The latter is a function of color, and, since we
will analyze the stars in color bins, the weights reduce to
$10^{-0.4m}$ where $m$ is the apparent magnitude in some passband.
Such a technique greatly increases the importance of the faintest and
most distant stars in the sample, i.e.\ those objects that are most
likely affected by extinction. Experimenting with such a scheme, we
found no significant reduction in the estimated uncertainties.

Another method to reduce the errors is to exclude high-proper-motion
stars.  For example, $\kappa$-$\sigma$-clipping excludes stars whose
$\o{\mu}_{\ell^\star}$ or $\mu_b$ deviates from the mean at their
longitude by more than $\kappa$ times the dispersion $\sigma$. This
approach is somewhat dangerous as it uses kinematic information itself
in a study aimed at kinematic quantities.

The most conservative technique is to exclude just bright stars, since
these are most likely very near and their proper motions are dominated
by random velocities.

\subsection{Correction for Mode Mixing} \label{sec:tech-mix}
Having obtained the raw Fourier coefficients
$(c_\ell,s_\ell,c_b,s_b)_m$, we might want to correct them for the
mode mixing described in \Sec{mix}, i.e.\ extract the Fourier
coefficients that would have been measured without longitudinal
variations in the mean parallax. Equating \eq{mu-fit} to \eq{mix-mu}
one gets a linear relation between the raw Fourier coefficients
$(c_\ell,s_\ell,c_b,s_b)_m$ and the unknowns\footnote{We will not try
  to solve for $\tilde{K}$ separately for two reasons.  First, the
  only distinction between $\o{\mu}_{W0}$ and $\sin b\,\tilde{K}$ in
  \eqn{mix-mub} is due to the mode-mixing terms, which are usually
  small.  Second, in almost all cases $\sin b\,\tilde{K}$ is much
  smaller than $\o{\mu}_{W0}$ and neglecting it does not introduce a
  significant error.}
\begin{equation}
  \B{\xi}\equiv(\o{\mu}_{U0},\,\o{\mu}_{V0},\,\o{\mu}_{W0},\,\tilde{A},\,\tilde{B},\,\tilde{C},\,\ldots),
\end{equation}
where the dots stands for possible higher-order terms.

\subsubsection{At Low Latitudes} \label{sec:tech-mix-low}
As pointed out in \Sec{mix}, at low latitudes $\o{\sin b}\simeq0$,
i.e.\ the Galactic-rotation induced higher-order terms in $\o{\mu}_b$
vanish, and we can use $\o{\mu}_b$ to directly measure the mode-mixing
coefficients $\B{\alpha}\equiv( c_\pi,s_\pi)_m$, see Equations
\eq{mean-pi} and \eq{mix-mub-low}. In \Sec{analysis} below, we will
determine the unknowns $\B{\xi}$ up to order $m= 4$, and need the
mode-mixing coefficients and hence the raw Fourier coefficients of
$\o{\mu}_b$ up to order $m=5$. Thus, we have 20 (non-linear) equations
(9 from $\o{\mu}_{\ell^\star}$ and 11 from $\o{\mu}_b$) for 20
unknowns (10 from $\B{\xi}$ and 10 from $\B{\alpha}$).

\subsubsection{At High Latitudes} \label{sec:tech-mix-high}
At high latitudes, we cannot estimate $\B{\alpha}$ directly from the
proper motions. However, if one is only interested in the low-order
coefficients (up to $m=2$ in \eqn{mix-mu}, the situation is not too
bad, as there are 10 equations for 6 unknowns in $\B{\xi}$ and another
6 in $\B{\alpha}$. Thus, even though the equations have no unique
solution, there is no complete freedom in the possible mode mixing. In
order to get a unique answer, one may postulate that mode mixing is
minimal, i.e.\ seek the smallest $\B{\alpha}$ that solves the
equations.

Note that in the analysis presented in \Sec{analysis} we will restrict
ourselves to low latitudes and use the procedure outlined in
\Sec{tech-mix-low}.

\subsection{Numerics and Error Analysis} \label{sec:tech-num}
\subsubsection{The Raw Fourier Coefficients} \label{sec:tech-num-raw}
Let us, for convenience, denote the vector of Fourier coefficients
$(c,s)_m$ by $\B{c}$ and that of the harmonic basis functions by
$\B{y}$. Then, we determine $\B{c}$ by minimizing the sample average
$\ave{(\mu - \B{c}\cdot\B{y})^2}$ giving
\begin{equation}
        \B{c} = \ave{\B{y}\otimes\B{y}}^{-1} \cdot \ave{\mu\B{y}}.
\end{equation}
The variance ${\bf V}_{\B{c}}$ is computed via error-propagation from
the variance of $\avet{\mu\B{y}}$, which is given by the standard
formula
\begin{equation}
  \mathbf{V}_{\avet{\mu\B{y}}} = \frac{\ave{\mu\B{y}\otimes\mu\B{y}} 
    - \ave{\mu\B{y}}\otimes\ave{\mu\B{y}}}{ N-1}.
\end{equation}

\subsubsection{Correction for Mode Mixing} \label{sec:tech-num-mix}
The $\B{c}_i$, $i=\ell,b$, are linearly related to our desired quantity
$\B{\xi}$,
\begin{equation} \label{eq:mm-eq}
  \B{c}_i = {\bf M}_i \cdot \B{\xi},
\end{equation}
where the matrices ${\bf M}_i$ are given through \eqn{mix-mu} and
depend linearly on the mode-mixing coefficients $\B{\alpha}$. Let
\begin{equation} \label{eq:chi-mu}
  \chi^2_{\B{\mu}} = \sum_{i=\ell,b} 
  (\B{c}_i - \mathbf{M}_i\cdot\B{\xi})^T\, 
  \mathbf{V}_{\B{c}_i}^{-1}\,
  (\B{c}_i - \mathbf{M}_i\cdot\B{\xi})
\end{equation}
be the error-weighted deviation from this relation. Solving \eq{mm-eq}
is equivalent to minimizing \eq{chi-mu} with respect to $\B{\xi}$ and
$\B{\alpha}$. The variance ${\bf V}_{\B{\xi}}$ is then given by the
$\B{\xi}$-part of twice the inverse of the Hessian matrix of
$\chi^2_{\B{\mu}}$.
\section{Analysis of the Proper Motion Catalogs} \label{sec:analysis}
The coordinate system defined by the Hipparcos Catalogue
\citep{HipTyc} is currently the best realization of an inertial
reference frame \citep{HippFrame97}. It is thus expedient to
re-determine the Oort constants using proper motion data based on this
reference frame. Unfortunately, the Hipparcos Catalogue itself is not
a kinematically unbiased catalog, so that special care has to be taken
when inferring kinematic properties from it \citep{Venice1}. A
kinematically unbiased subsample extracted from the Hipparcos
Catalogue by DB98 contains about 14\,000 stars with a maximum and
average distance of about 100 and 80 parsec. Using our estimate from
\Sec{accur} for the accuracy of the Oort constants, we expect for this
sample an uncertainty of the Oort constants of about 4\,\kmskpc, which is
unacceptably large.

In order to get more accurate estimates for the Oort constants, one
needs deeper and/or larger samples.  In the present paper we will
employ the ACT Catalog \citep{ACT}, which is essentially magnitude
limited, and hence kinematically unbiased.  The ACT Catalog is a
combination of the Astrographic Catalog \citep{hE74,ACR88,AC2000} with
a median epoch near 1904, and the Tycho Catalogue \citep{HipTyc},
which has a limiting magnitude of about $V=11.5$. It contains, for
nearly $10^6$ stars, two color photometry\footnote{ We use the
  photometric system as defined in the ACT catalog, which in turn was
  taken from the Tycho Catalogue.  That is to say, when we speak about
  apparent magnitudes we refer to ``Tycho'' magnitudes.  These Tycho
  magnitudes have been used to calculate an approximate color in the
  Johnson system [(\bv)$_{\rm J}]$.  Note that these color
  transformations depends somewhat on spectral type and luminosity
  class. Hence, by necessity, the ACT colors are only approximate,
  especially in the region where there is a transition between
  predominantly main-sequence stars to mainly giants (i.e., the
  sub-giant region). For details on the Tycho photometric system we
  refer to The Hipparcos and Tycho Catalogues \citep[][Chapter
  16]{HipTyc}.} and accurate astrometric data (positions and proper
motions) defined in the Hipparcos reference frame.

We have also analyzed the Tycho-2 catalog \citep{Tyc2a}. This catalog
entails a re-analysis of the data of Hipparcos' Tycho instrument, the
astrographic catalog and many other astrometric datasets
\citep{Tyc2b}. This catalog contains about twice as many stars as ACT
catalog, where the majority of the ``extra'' stars are fainter than
$V_T\sim10$, while its astrometry is equivalent to that of the ACT
catalog \citep{Tyc2a}. It might be tempting to use the larger Tycho-2
catalog instead of the ACT. However, a twice larger sample means a
roughly 40\% larger limiting distance, so that the distance-dependent
effects and higher-order terms may become important (see
\Sec{sys-depth}).  Instead we decided to use the ACT catalog, and
verify the results by a similar analysis of the Tycho-2 data over the
ACT magnitude range ($8.5 \le V_T \le 11.5$). Since the ACT and
Tycho-2 results are entirely consistent with each other, we will use
the term ``ACT/Tycho-2'' to describe our data in the remainder of this
paper.

Compared to the Hipparcos sub-sample, the sample-depth and the number
of stars is increased by factors of about ten and 70,
respectively. Estimating the expected accuracy for the Oort constants,
one realizes that we may even divide this sample into subsets and
still get reasonably accurate values for $\tilde{A}$, $\tilde{B}$, and
$\tilde{C}$.
\begin{figure*}[t]
   \ifpreprint
      \centerline{\resizebox{170mm}{!}{\includegraphics{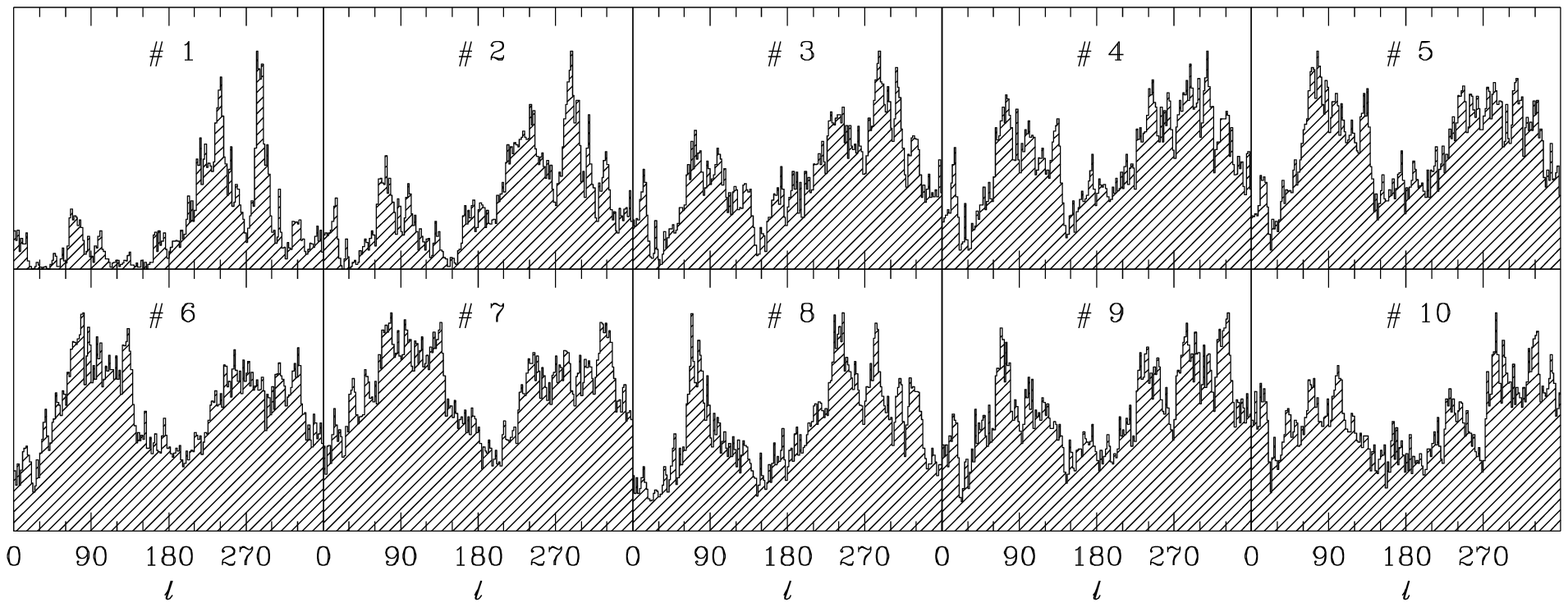}}}
   \else
      \centerline{\resizebox{160mm}{!}{\plotone{f3}}}
   \fi
   \input{ms_fc03}
\end{figure*}
%
\subsection{Longitudinal Inhomogeneities} \label{sec:nonuni}
As discussed in \Sec{measure}, spatial inhomogeneity and extinction
are expected to bias the Oort constants and to confuse them via mode
mixing.  We are able, though, to correct for the latter effect
provided we restrict ourselves to low latitudes.  For this reason,
since the stellar density is highest there, and because we are mainly
interested in the Oort constants in the Galactic plane, we restrict our
analysis to stars with $|\sin b|\le0.1$ ($|b|\la 5.74^\circ$).  After
further excluding stars brighter than $m_B=8.5$ (\Sec{weight}), we are
left with 192\,546 stars in the color range $\bv\in[-0.2,4]$.
\ifpreprint

\begin{table}[t]
  \begin{center}
    \scshape
    \refstepcounter{table} \label{tab:bins}
    Table \arabic{table} \\
    The Color Bins
  \end{center}
  \footnotesize
  \begin{tabular*}{\columnwidth}{@{\extracolsep{\fill}}
     r@{\hspace{-0.3em}}
     l@{\hspace{ 0.2em}}
     r@{ -\hspace{-0.1em}} 
     l@{\hspace{ 0.3em}}   
     r@{\hspace{-0.2em}}   
     c@{\hspace{-0.3em}}   
     c@{\hspace{-1.0em}}   
     c@{\hspace{ 0.0em}}   
     c@{}                  
     r@{}                  
    }
    \hline \hline
    \\[-1.5ex]
    \multicolumn{2}{r@{\hspace{ 0.0em}}}{\#} &
    \multicolumn{2}{c@{\hspace{ 0.0em}}}{$\bv$} &
    \multicolumn{1}{c@{\hspace{ 0.0em}}}{$N$} &
    \multicolumn{1}{c@{\hspace{ 0.0em}}}{$\delta n(\ell)$} &
    \multicolumn{1}{c@{\hspace{ 0.0em}}}{$\frac{1}{\o{\pi_W}}$} &
    \multicolumn{1}{c@{\hspace{ 0.0em}}}{$(\bv)_0$} &
    \multicolumn{1}{c@{\hspace{ 0.0em}}}{$\tau_{\mathrm{MS}}$} &
    \multicolumn{1}{c@{\hspace{ 0.0em}}}{$f_Y$}
    \\[0.3ex]
    \hline
    \\[-2ex]
    1& & --0.20 & 0.00 &  7,843 & 0.272 & 0.652 &--0.31 & 0.01  & 100 \\
    1&a& --0.10 & 0.05 & 12,931 & 0.236 & 0.732 &--0.26 & 0.02  & 100 \\
    2& &   0.00 & 0.10 & 14,345 & 0.180 & 0.724 &--0.18 & 0.03  & 100 \\
    2&a&   0.05 & 0.15 & 16,490 & 0.139 & 0.703 &--0.13 & 0.07  & 100 \\
    3& &   0.10 & 0.20 & 17,708 & 0.120 & 0.683 &--0.07 & 0.18  & 100 \\
    3&a&   0.15 & 0.25 & 18,890 & 0.111 & 0.646 &--0.01 & 0.38  & 100 \\
    4& &   0.20 & 0.30 & 19,097 & 0.095 & 0.583 &  0.06 & 0.59  & 100 \\
    4&a&   0.25 & 0.36 & 21,211 & 0.088 & 0.539 &  0.13 & 0.77  & 100 \\
    5& &   0.30 & 0.42 & 23,611 & 0.085 & 0.488 &  0.20 & 1.07  & 100 \\
    5&a&   0.36 & 0.50 & 27,824 & 0.071 & 0.422 &  0.29 & 1.64  &  91 \\
    6& &   0.42 & 0.55 & 25,815 & 0.059 & 0.375 &  0.36 & 2.61  &  57 \\
    6&a&   0.50 & 0.70 & 30,619 & 0.041 & 0.301 &  0.50 & 5.55  &  27 \\
    7& &   0.55 & 0.85 & 31,234 & 0.043 & 0.279 &  0.61 & 8.41  &  18 \\[2ex]
    7&a&   0.70 & 0.97 & 17,987 & 0.089 & 0.309 &  0.74 &\nodata&  64 \\
    8& &   0.85 & 1.10 & 18,056 & 0.137 & 0.475 &  0.82 &\nodata& 100 \\
    8&a&   0.97 & 1.20 & 17,993 & 0.122 & 0.578 &  0.90 &\nodata&  96 \\
    9& &   1.10 & 1.35 & 16,072 & 0.083 & 0.652 &  1.01 &\nodata&  58 \\
    9&a&   1.20 & 1.50 & 14,198 & 0.065 & 0.747 &  1.11 &\nodata&  40 \\
    10&&   1.35 & 1.90 & 12,797 & 0.046 & 0.854 &  1.35 &\nodata&  22 \\[0.3ex]
    \hline
  \end{tabular*}\par \vspace{1ex}
  \hspace{1em} {\scshape Note.---}
    The boundaries of the (observed) \bv\ color; number $N$ of stars
    fainter than $V_T=8.5$; relative overabundance in $n(\ell)$ (see
    \Sec{nonuni}); Average inverse parallax [in kpc, see \eqn{pi-w}];
    our best estimate of the intrinsic color; the main-sequence
    lifetime $\tau_{\mathrm{MS}}$ (in Gyr; bins 7a--10 comprise non-MS
    stars); and the percentage $f_Y$ of stars younger than 1.5 Gyr.
    For the main-sequence stars, we estimate $f_Y$ from the MS
    lifetime and a constant SFR, for the post-MS star, we employ the
    $\delta n(\ell)$ method, see \Sec{Sample_properties} for details.

\end{table}

\else

\begin{deluxetable}{r@{}lr@{ - }lrccccr}
  \tabletypesize{\footnotesize}
  \tablewidth{0pt}
  \tablecaption{The Color Bins\label{tab:bins}}
  \tabletypesize{\footnotesize}  \tablehead{
    \multicolumn{2}{c}{\#} &
    \multicolumn{2}{c}{$\bv$} &
    \multicolumn{1}{c}{$N$} &
    \multicolumn{1}{c}{$\delta n(\ell)$} &
    \multicolumn{1}{c}{$\frac{1}{\o{\pi_W}}$} &
    \multicolumn{1}{c}{$(\bv)_0$} &
    \multicolumn{1}{c}{$\tau_{\mathrm{MS}}$} &
    \multicolumn{1}{c}{$f_Y$}
    }
  \tablecolumns{10}
  \startdata
  1& & --0.20 & 0.00 &  7,843 & 0.272 & 0.652 &--0.31 & 0.01  & 100 \\
  1&a& --0.10 & 0.05 & 12,931 & 0.236 & 0.732 &--0.26 & 0.02  & 100 \\
  2& &   0.00 & 0.10 & 14,345 & 0.180 & 0.724 &--0.18 & 0.03  & 100 \\
  2&a&   0.05 & 0.15 & 16,490 & 0.139 & 0.703 &--0.13 & 0.07  & 100 \\
  3& &   0.10 & 0.20 & 17,708 & 0.120 & 0.683 &--0.07 & 0.18  & 100 \\
  3&a&   0.15 & 0.25 & 18,890 & 0.111 & 0.646 &--0.01 & 0.38  & 100 \\
  4& &   0.20 & 0.30 & 19,097 & 0.095 & 0.583 &  0.06 & 0.59  & 100 \\
  4&a&   0.25 & 0.36 & 21,211 & 0.088 & 0.539 &  0.13 & 0.77  & 100 \\
  5& &   0.30 & 0.42 & 23,611 & 0.085 & 0.488 &  0.20 & 1.07  & 100 \\
  5&a&   0.36 & 0.50 & 27,824 & 0.071 & 0.422 &  0.29 & 1.64  &  91 \\
  6& &   0.42 & 0.55 & 25,815 & 0.059 & 0.375 &  0.36 & 2.61  &  57 \\
  6&a&   0.50 & 0.70 & 30,619 & 0.041 & 0.301 &  0.50 & 5.55  &  27 \\
  7& &   0.55 & 0.85 & 31,234 & 0.043 & 0.279 &  0.61 & 8.41  &  18 \\[2ex]
  7&a&   0.70 & 0.97 & 17,987 & 0.089 & 0.309 &  0.74 &\nodata&  64 \\
  8& &   0.85 & 1.10 & 18,056 & 0.137 & 0.475 &  0.82 &\nodata& 100 \\
  8&a&   0.97 & 1.20 & 17,993 & 0.122 & 0.578 &  0.90 &\nodata&  96 \\
  9& &   1.10 & 1.35 & 16,072 & 0.083 & 0.652 &  1.01 &\nodata&  58 \\
  9&a&   1.20 & 1.50 & 14,198 & 0.065 & 0.747 &  1.11 &\nodata&  40 \\
  10&&   1.35 & 1.90 & 12,797 & 0.046 & 0.854 &  1.35 &\nodata&  22 \\ 
  \enddata
  \ifpreprint\vspace{-4.5ex}\fi
  \tablecomments{The boundaries of the (observed) \bv\ color; number
    $N$ of stars fainter than $V_T=8.5$; relative overabundance
    in $n(\ell)$ (see \Sec{nonuni}); Average inverse parallax [in kpc,
    see \eqn{pi-w}]; our best estimate of the intrinsic
    color; the main-sequence lifetime $\tau_{\mathrm{MS}}$ (in Gyr;
    bins 7a--10 comprise non-MS stars); and the percentage $f_Y$ of
    stars younger than 1.5 Gyr.  For the main-sequence stars, we
    estimate $f_Y$ from the MS lifetime and a constant SFR, for the
    post-MS star, we employ the $\delta n(\ell)$ method, see
    \Sec{Sample_properties} for details.}
\end{deluxetable}

\fi

In order to get a feeling for the importance of the spatial
inhomogeneities, we binned the stars into 60 bins in $\bv$ and
considered their longitudinal frequency, $n(\ell)$.  \Fig{nl} plots
$n(\ell)$ for the color bins 1 to 10 (see \Tab{bins}) used in the
analysis below. Most prominent in the bluest color bin (\#1) are three
peaks near $\ell\simeq75^\circ,\,240^\circ$, and $285^\circ$, which
presumably are caused by stars in the Sagittarius-Carina
($\ell=75\arcdeg$, $285\arcdeg $) and the Orion-Cygnus
($\ell=240\arcdeg$) spiral-arms. In order to quantify how important
these peaks are, we measured the relative number $\delta n(\ell)$ of
stars in three windows of $30^\circ$ width that were centered on the
three peaks, and subtracted $1/4$, the expectation value for a uniform
distribution.  \Fig{nonuni} plots this measure versus the mean color
for 60 narrow color bins (open), as well as the 19 bins used in the
analysis below (solid). The importance of the peaks is largest at the
blue end, where the stars are very young and not mixed or settled into
equilibrium.  Moreover, these stars are very bright and can be seen
out to a few \kpc, such that patchy extinction can best contribute to
the apparent non-uniformity.  Along the main sequence (MS), where the
stars get older, hence better mixed, and fainter, $n(\ell)$ steadily
becomes more uniform until $\bv\simeq0.7$.  Then the peaks become
important again near $\bv\approx1$, where the red clump dominates the
sample, while the giants at $\bv\ga1.2$ are nearly as uniformly
distributed as are the dwarfs near $\bv\simeq0.7$.

The red-most of the 60 narrow bins at $\bv=1.95$ is somewhat odd, as
it has less stars in those peaks than even for uniform $n(\ell)$.  The
likely reason is that these are stars subject to severe extinction,
which made them appear much redder than they really are (there are no
stars with intrinsic $\bv>2$), and restricts them to regions where
high extinction has diminished the numbers in the other color bins.

\begin{figure}[h]
   \ifpreprint
      \centerline{\resizebox{85mm}{!}{\includegraphics{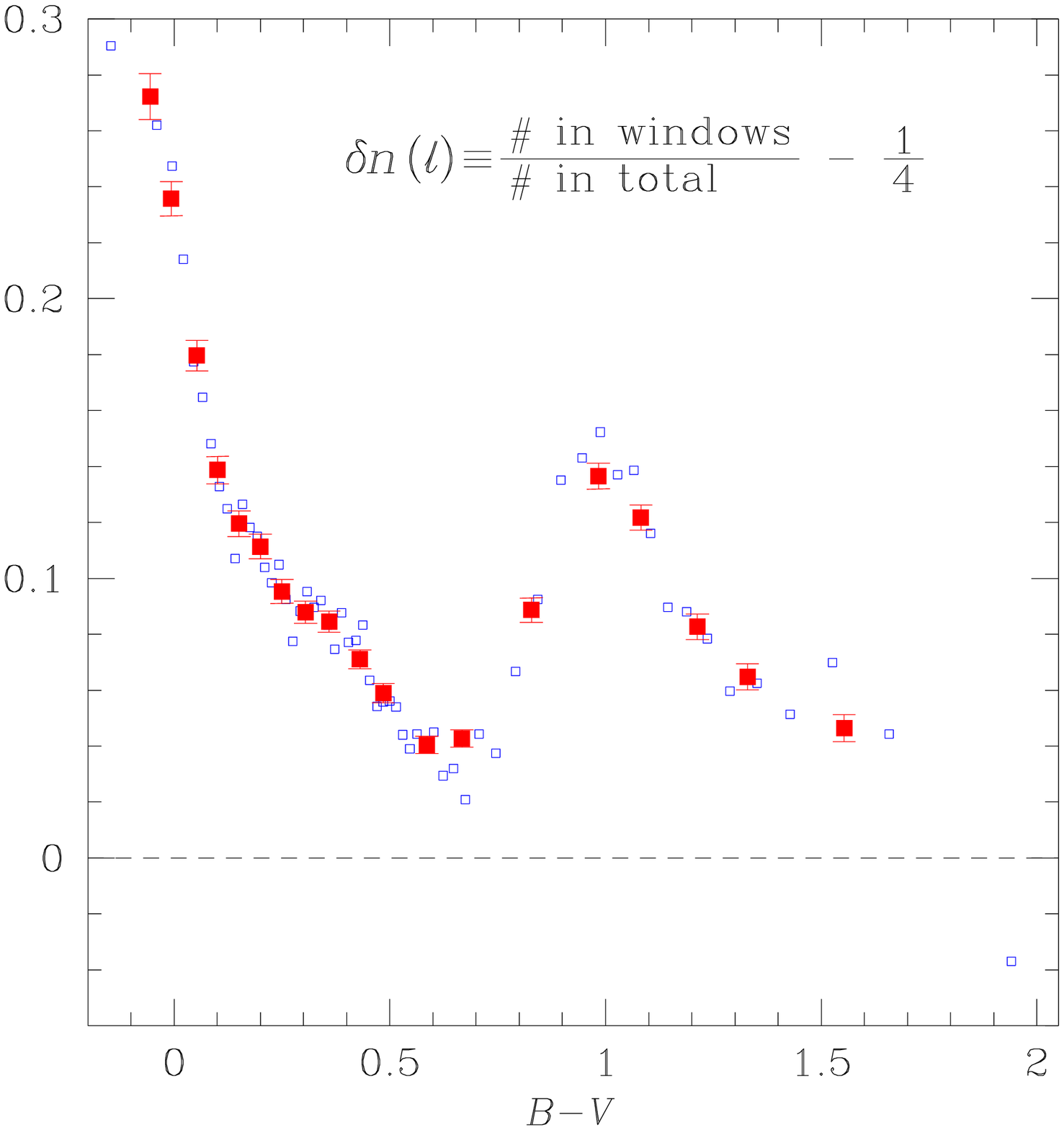}}}
   \else
      \centerline{\resizebox{70mm}{!}{\plotone{f4}}}
   \fi
   \input{ms_fc04}
\end{figure}
%
\subsection{Data Analysis}
\subsubsection{Binning and Analyzing the Data}
Before the analysis in terms of the Oort constants, we split the data it
into ten color bins, labeled 1 to 10.  These bins have been chosen to
group together stars with similar $n(\ell)$.  Small bins had to be
avoided, in particular at red colors, to yield reasonable errors for
the Oort constants -- large errors would render differences between
the results of adjacent bins insignificant.  In addition to these ten
distinct bins, we consider 9 bins of similar size (labeled 1a to 9a)
whose stars are taken half from each of the nearest primary bins.  The
color limits of all the 19 bins are given in \Tab{bins}, which also
lists the numbers of stars fainter than $V_T=8.5$ contained in each
bin.

To reduce the errors, we also analyzed the data after applying a
modest $\kappa$-$\sigma$-clipping with $\kappa=5$ and excluding not
more than 5\% of the stars in each of 20 longitudinal bins (the only
usage ever of such bins).  The raw Fourier coefficients for
$\o{\mu}_{\ell^\star}$ and $\o{\mu}_b$ and their errors were estimated
up to $m_{\rm max}=32$ as outlined in \Sec{tech-num-raw} and are
tabulated up to $m=5$ in Tables \ref{tab:raw-ml} and \ref{tab:raw-mb}
for all 19 color bins. Note that if there were no mode mixing,
$\tilde{A}$, $\tilde{B}$, and $\tilde{C}$ could be read of
\Tab{raw-ml} as $c_2$, $c_0$, and $-s_2$.

\begin{deluxetable}{r@{}l@{\quad} 
    r@{${\pm}$}l                                    
    r@{${\pm}$}l@{$\;\;$}r@{${\pm}$}l               
    r@{${\pm}$}l@{$\;\;$}r@{${\pm}$}l               
    r@{${\pm}$}l@{$\;\;$}r@{${\pm}$}l               
    r@{${\pm}$}l@{$\;\;$}r@{${\pm}$}l               
    r@{${\pm}$}l@{$\;\;$}r@{${\pm}$}l}              
  \ifpreprint
     \tabletypesize{\footnotesize}
  \else
     \tabletypesize{\scriptsize}
  \fi
  \tablewidth{0pt}
  \tablecaption{Raw Fourier Coefficients for $\o{\mu}_{\ell^\star}$
    \label{tab:raw-ml}}
  \tablehead{
    \multicolumn{2}{c}{bin}   & \multicolumn{2}{c}{$c_0$} &
    \multicolumn{2}{c}{$c_1$} & \multicolumn{2}{c}{$s_1$} &
    \multicolumn{2}{c}{$c_2$} & \multicolumn{2}{c}{$s_2$} &
    \multicolumn{2}{c}{$c_3$} & \multicolumn{2}{c}{$s_3$} &
    \multicolumn{2}{c}{$c_4$} & \multicolumn{2}{c}{$s_4$} &
    \multicolumn{2}{c}{$c_5$} & \multicolumn{2}{c}{$s_5$}
    }
  \tablecolumns{24}
  \startdata
 1&  &--11.8&1.1 &--13.9&1.7 &   9.8&1.5 &  12.4&1.5 & --0.2&1.7 & --0.1&1.4 & --5.4&1.8 &   0.7&1.7 &   0.0&1.6 & --0.1&1.7 &   1.4&1.6\\
 1&a &--12.0&0.6 &--14.6&1.0 &  12.1&0.8 &  11.8&0.8 &   1.1&0.9 &   0.4&0.7 & --4.8&1.0 &   0.4&0.9 &   1.4&0.9 &   0.1&0.9 &   1.3&0.9\\
 2&  &--11.9&0.4 &--15.4&0.6 &  13.3&0.5 &  13.4&0.5 &   1.7&0.6 &   0.9&0.4 & --4.4&0.7 &   0.1&0.5 &   1.2&0.6 & --0.7&0.5 &   0.8&0.6\\
 2&a &--11.9&0.3 &--17.2&0.5 &  14.0&0.4 &  14.2&0.4 &   1.8&0.5 &   0.7&0.4 & --3.4&0.5 & --0.1&0.5 &   0.5&0.5 &   0.0&0.5 &   0.4&0.5\\
 3&  &--11.9&0.4 &--17.9&0.6 &  14.0&0.5 &  14.7&0.5 &   1.8&0.6 & --0.0&0.5 & --3.9&0.7 &   0.2&0.5 &   0.8&0.7 &   0.6&0.6 & --0.0&0.6\\
 3&a &--11.8&0.4 &--19.3&0.6 &  14.3&0.5 &  15.6&0.5 &   1.3&0.6 & --1.0&0.5 & --3.6&0.6 &   0.8&0.6 &   0.1&0.6 &   1.0&0.6 &   0.6&0.5\\
 4&  &--12.0&0.4 &--21.9&0.6 &  15.8&0.6 &  14.8&0.6 &   0.9&0.6 & --0.7&0.6 & --3.2&0.6 &   0.6&0.6 & --0.5&0.6 &   1.2&0.7 &   0.8&0.6\\
 4&a &--12.1&0.5 &--25.6&0.8 &  17.4&0.6 &  14.7&0.7 &   1.4&0.7 & --0.4&0.7 & --3.3&0.7 &   0.9&0.6 & --0.0&0.8 &   0.6&0.7 & --0.2&0.7\\
 5&  &--12.4&0.6 &--28.8&0.9 &  20.3&0.7 &  15.7&0.8 &   1.6&0.8 & --1.6&0.8 & --3.2&0.9 &   1.5&0.8 &   0.1&0.8 &   0.4&0.9 & --0.8&0.8\\
 5&a &--11.7&0.6 &--36.2&0.9 &  24.5&0.7 &  18.2&0.9 &   1.6&0.8 & --1.9&0.8 & --3.4&0.8 & --0.3&0.8 & --0.4&0.8 & --0.8&0.9 & --1.0&0.8\\
 6&  &--10.7&0.7 &--43.9&1.0 &  26.5&0.9 &  18.8&1.0 &   2.0&0.9 & --1.9&1.0 & --2.8&1.0 & --0.1&1.0 &   0.5&0.9 & --1.7&1.0 & --0.6&0.9\\
 6&a & --9.6&0.9 &--63.7&1.3 &  30.9&1.1 &  17.7&1.3 &   3.0&1.2 & --3.7&1.2 & --1.8&1.3 &   1.7&1.2 &   0.5&1.3 & --0.4&1.3 &   0.8&1.3\\
 7&  & --9.6&1.0 &--75.7&1.5 &  32.1&1.3 &  16.4&1.5 &   1.9&1.4 & --5.2&1.4 & --2.4&1.4 &   1.3&1.4 & --1.1&1.5 & --0.5&1.4 &   0.9&1.4\\
 7&a & --9.8&1.4 &--81.1&2.2 &  28.9&1.7 &  11.6&2.2 & --5.7&1.9 & --7.7&2.1 &   0.8&2.0 &   0.5&2.1 & --3.5&2.0 & --2.1&2.0 & --0.1&2.1\\
 8&  & --9.8&1.4 &--60.2&2.2 &  23.5&1.7 &  13.7&1.9 & --9.4&2.1 & --2.6&1.8 &   4.7&2.2 & --0.8&2.0 & --5.3&2.0 &   1.8&2.0 &   1.9&2.0\\
 8&a &--12.2&1.1 &--45.9&1.8 &  20.1&1.4 &  14.2&1.5 & --5.4&1.8 &   0.2&1.4 &   1.3&1.9 &   0.1&1.6 & --3.8&1.7 &   3.7&1.6 &   2.1&1.8\\
 9&  &--13.6&1.0 &--40.0&1.6 &  15.3&1.4 &  13.7&1.3 & --0.5&1.7 & --0.8&1.3 & --0.6&1.7 &   1.6&1.6 & --0.8&1.4 &   0.7&1.6 &   0.4&1.4\\
 9&a &--12.8&1.1 &--38.6&1.7 &  13.9&1.5 &  14.2&1.3 & --1.9&1.9 &   0.0&1.5 &   1.5&1.8 &   1.0&1.8 & --2.0&1.4 &   0.5&1.9 &   1.1&1.4\\
10&  &--12.2&0.9 &--30.5&1.3 &  13.5&1.2 &  13.8&1.0 & --3.0&1.5 &   1.2&1.2 &   1.7&1.4 & --0.9&1.4 & --1.4&1.3 &   0.4&1.3 &   1.7&1.3\\

  \cutinhead{with $\kappa$-$\sigma$-clipping}
 1&  &--13.3&0.6 &--15.3&0.9 &  11.8&0.9 &  13.1&0.7 &   0.5&1.0 &   0.3&0.8 & --4.8&1.0 &   2.3&0.9 &   0.7&1.0 &   1.3&1.0 &   2.2&0.8\\
 1&a &--12.5&0.4 &--15.2&0.6 &  13.0&0.6 &  13.2&0.5 &   1.5&0.7 &   0.4&0.5 & --4.2&0.7 &   1.0&0.6 &   1.0&0.6 &   0.1&0.6 &   1.8&0.5\\
 2&  &--11.9&0.3 &--15.8&0.4 &  13.6&0.4 &  13.5&0.4 &   1.7&0.5 &   1.2&0.4 & --3.9&0.5 &   0.1&0.5 &   1.1&0.4 & --0.1&0.5 &   0.9&0.4\\
 2&a &--11.9&0.2 &--17.0&0.4 &  13.8&0.4 &  13.7&0.3 &   1.7&0.4 &   1.1&0.3 & --3.4&0.4 &   0.2&0.4 &   0.9&0.4 &   0.3&0.4 &   0.6&0.4\\
 3&  &--12.0&0.2 &--18.2&0.4 &  13.8&0.3 &  14.5&0.3 &   1.5&0.4 &   0.2&0.3 & --3.4&0.4 &   0.3&0.4 &   0.2&0.4 &   0.6&0.4 &   0.6&0.4\\
 3&a &--11.9&0.2 &--19.5&0.3 &  14.1&0.3 &  14.8&0.4 &   1.7&0.4 & --0.8&0.3 & --3.1&0.4 &   0.2&0.4 & --0.3&0.4 &   0.4&0.4 &   0.9&0.4\\
 4&  &--11.8&0.3 &--20.9&0.4 &  16.1&0.4 &  14.6&0.4 &   2.4&0.4 & --1.1&0.4 & --2.5&0.4 &   0.1&0.4 & --0.3&0.4 &   0.0&0.4 &   0.9&0.4\\
 4&a &--11.5&0.3 &--23.1&0.4 &  18.1&0.4 &  15.4&0.4 &   2.5&0.4 & --0.5&0.4 & --2.7&0.4 &   1.0&0.4 &   0.0&0.4 & --0.2&0.4 &   0.3&0.4\\
 5&  &--11.0&0.3 &--25.6&0.5 &  20.3&0.4 &  16.2&0.5 &   2.7&0.5 & --1.0&0.4 & --2.7&0.5 &   0.8&0.5 & --0.1&0.5 & --0.1&0.5 &   0.1&0.5\\
 5&a &--10.9&0.4 &--31.9&0.5 &  23.7&0.5 &  18.0&0.6 &   3.0&0.5 & --1.7&0.5 & --2.5&0.6 &   0.2&0.5 & --0.1&0.6 & --0.5&0.6 & --0.1&0.5\\
 6&  &--10.7&0.5 &--38.7&0.7 &  25.1&0.7 &  18.6&0.7 &   2.6&0.7 & --1.9&0.7 & --2.1&0.7 &   0.4&0.7 &   1.4&0.7 & --0.6&0.7 & --0.8&0.7\\
 6&a & --9.9&0.6 &--53.8&0.9 &  28.3&0.9 &  17.6&1.0 &   3.2&0.9 & --3.7&0.9 & --2.2&0.9 &   1.3&0.9 &   0.6&0.9 & --1.6&0.9 & --0.0&0.9\\
 7&  & --9.6&0.7 &--63.5&1.0 &  30.4&1.0 &  18.1&1.1 &   1.5&1.0 & --3.5&1.0 & --1.3&1.0 &   1.7&1.0 & --0.8&1.1 & --0.8&1.0 & --0.5&1.0\\
 7&a &--10.1&1.0 &--66.2&1.5 &  27.1&1.3 &  13.2&1.5 & --7.4&1.5 & --3.9&1.4 &   1.7&1.5 &   0.5&1.4 & --4.3&1.5 &   0.2&1.4 & --2.3&1.5\\
 8&  & --9.5&0.7 &--42.2&1.2 &  19.7&0.9 &  15.7&1.0 & --5.6&1.2 & --0.3&1.0 &   3.8&1.2 & --2.2&1.1 & --3.8&1.1 &   2.6&1.1 &   1.1&1.0\\
 8&a &--11.3&0.5 &--34.2&0.8 &  16.3&0.7 &  16.4&0.7 & --0.2&0.8 &   0.2&0.7 &   1.2&0.9 & --0.1&0.8 & --1.1&0.8 &   1.3&0.8 &   0.8&0.8\\
 9&  &--12.7&0.5 &--33.0&0.7 &  14.3&0.6 &  15.4&0.7 &   2.0&0.7 & --0.5&0.7 &   0.6&0.7 &   0.9&0.7 &   0.4&0.7 & --0.1&0.7 &   0.6&0.7\\
 9&a &--12.6&0.4 &--31.6&0.6 &  13.9&0.6 &  14.4&0.7 &   1.0&0.7 & --0.0&0.6 &   1.4&0.7 &   0.3&0.7 &   0.1&0.7 & --0.9&0.7 &   1.2&0.7\\
10&  &--12.5&0.4 &--27.4&0.6 &  13.0&0.6 &  14.2&0.6 & --1.0&0.6 &   0.9&0.6 &   2.1&0.6 & --1.2&0.6 &   0.3&0.6 & --0.9&0.6 &   1.4&0.6\\

  \enddata
\end{deluxetable}

\begin{deluxetable}{r@{}l@{\quad}
    r@{${\pm}$}l                                    
    r@{${\pm}$}l@{$\;\;$}r@{${\pm}$}l               
    r@{${\pm}$}l@{$\;\;$}r@{${\pm}$}l               
    r@{${\pm}$}l@{$\;\;$}r@{${\pm}$}l               
    r@{${\pm}$}l@{$\;\;$}r@{${\pm}$}l               
    r@{${\pm}$}l@{$\;\;$}r@{${\pm}$}l}              
  \ifpreprint
     \tabletypesize{\footnotesize}
  \else
     \tabletypesize{\scriptsize}
  \fi
  \tablewidth{0pt}
  \tablecaption{Raw Fourier Coefficients for $\o{\mu}_b$ \label{tab:raw-mb}}
  \tablehead{
    \multicolumn{2}{c}{bin}   & \multicolumn{2}{c}{$c_0$} &
    \multicolumn{2}{c}{$c_1$} & \multicolumn{2}{c}{$s_1$} &
    \multicolumn{2}{c}{$c_2$} & \multicolumn{2}{c}{$s_2$} &
    \multicolumn{2}{c}{$c_3$} & \multicolumn{2}{c}{$s_3$} &
    \multicolumn{2}{c}{$c_4$} & \multicolumn{2}{c}{$s_4$} &
    \multicolumn{2}{c}{$c_5$} & \multicolumn{2}{c}{$s_5$}
    }
  \tablecolumns{24}
  \startdata
 1&  &--11.9&1.7 & --1.3&2.7 & --1.4&2.0 & --0.0&1.5 & --2.9&3.0 &   2.9&1.2 & --2.4&3.1 &   1.0&2.4 & --0.7&2.4 &   4.2&3.1 & --0.0&1.4\\
 1&a &--11.3&1.0 &   0.3&1.6 & --0.7&1.1 & --0.1&0.9 & --1.6&1.7 &   2.9&0.7 & --2.7&1.8 &   0.4&1.3 &   0.6&1.4 &   1.2&1.7 & --0.7&0.8\\
 2&  &--10.0&0.3 &   1.9&0.5 &   0.2&0.4 &   0.5&0.4 & --0.1&0.4 &   1.9&0.4 & --0.8&0.5 &   0.1&0.4 &   1.2&0.5 &   0.3&0.4 &   0.3&0.4\\
 2&a &--10.3&0.2 &   1.0&0.3 & --0.0&0.3 &   0.4&0.3 & --0.7&0.4 &   1.2&0.4 & --0.7&0.4 &   0.2&0.4 &   1.4&0.4 &   0.5&0.3 &   0.7&0.4\\
 3&  &--11.0&0.4 &   1.6&0.7 & --0.9&0.5 & --0.2&0.5 & --0.9&0.7 &   1.1&0.4 & --1.4&0.7 & --0.0&0.4 &   2.5&0.7 &   0.5&0.6 &   0.2&0.6\\
 3&a &--11.2&0.4 &   2.3&0.6 & --0.8&0.5 & --1.0&0.5 & --0.8&0.6 &   0.7&0.4 & --1.1&0.6 & --0.2&0.5 &   2.5&0.6 &   0.1&0.6 &   0.6&0.5\\
 4&  &--11.9&0.3 &   1.9&0.5 & --0.0&0.4 & --0.4&0.5 & --1.7&0.5 &   1.0&0.5 & --0.8&0.5 & --0.2&0.4 &   1.4&0.5 &   0.2&0.5 &   0.5&0.5\\
 4&a &--13.4&0.3 &   2.4&0.6 &   0.2&0.4 & --0.0&0.6 & --1.7&0.5 &   2.4&0.5 & --0.8&0.5 &   0.0&0.5 &   2.0&0.6 &   0.6&0.5 &   0.8&0.6\\
 5&  &--15.2&0.4 &   3.2&0.6 & --0.6&0.4 & --0.1&0.6 & --1.4&0.5 &   2.7&0.6 & --0.6&0.5 & --0.0&0.5 &   2.6&0.6 &   0.0&0.5 &   1.8&0.6\\
 5&a &--18.2&0.4 &   3.7&0.6 & --0.4&0.4 & --0.8&0.6 & --1.5&0.5 &   1.8&0.6 & --0.7&0.5 & --0.6&0.5 &   1.7&0.5 &   0.8&0.5 &   1.4&0.5\\
 6&  &--21.1&0.5 &   3.0&0.8 & --0.5&0.5 & --2.2&0.8 & --1.9&0.6 &   1.6&0.7 & --0.8&0.7 & --0.4&0.7 &   1.0&0.7 &   1.0&0.7 &   1.0&0.7\\
 6&a &--25.9&0.6 &   1.5&1.0 & --1.9&0.7 & --3.6&0.9 & --2.9&0.8 &   0.6&0.8 & --2.2&0.9 & --0.1&0.8 & --0.4&0.9 & --0.2&0.8 & --0.7&0.9\\
 7&  &--28.2&0.6 &   1.2&1.1 & --2.3&0.8 & --4.7&1.0 & --3.7&0.9 &   0.5&0.9 & --1.8&0.9 & --1.2&0.9 & --0.9&0.9 & --0.5&0.9 &   0.2&0.9\\
 7&a &--28.4&1.0 &   1.2&1.7 & --4.3&1.1 & --8.1&1.5 & --3.8&1.3 &   0.0&1.5 & --1.6&1.3 & --1.7&1.4 & --1.1&1.3 &   0.5&1.4 &   2.8&1.4\\
 8&  &--21.1&0.9 &   2.3&1.6 & --5.2&1.0 & --5.6&1.3 & --1.3&1.3 & --0.8&1.3 & --3.9&1.4 &   1.3&1.3 & --0.6&1.3 &   1.9&1.3 & --0.2&1.3\\
 8&a &--15.4&0.7 &   1.8&1.3 & --4.0&0.8 & --3.3&1.0 & --0.0&1.1 & --0.2&0.9 & --3.0&1.2 & --0.0&1.1 &   1.1&1.1 &   1.5&1.1 & --0.7&1.0\\
 9&  &--12.5&0.6 &   0.5&0.9 & --3.9&0.7 & --2.9&0.7 & --1.1&0.9 & --0.3&0.8 & --2.9&0.9 & --1.0&0.8 &   0.8&0.8 &   2.5&0.8 &   0.3&0.9\\
 9&a &--11.7&0.6 &   0.9&1.0 & --4.1&0.8 & --2.7&0.7 & --1.0&1.1 & --0.4&0.8 & --3.0&1.0 & --0.3&1.0 & --0.1&0.9 &   1.2&1.0 & --0.4&0.9\\
10&  & --9.5&0.6 &   0.8&0.8 & --3.8&0.8 & --1.0&0.8 &   0.6&0.8 & --0.8&0.7 & --3.5&0.9 & --0.4&0.9 &   0.1&0.8 & --0.1&0.9 & --1.5&0.8\\

  \cutinhead{with $\kappa$-$\sigma$-clipping}
 1&  &--11.0&1.6 & --0.9&2.6 & --0.9&1.8 & --1.1&1.1 & --3.2&2.8 &   2.8&0.9 & --3.7&2.9 &   1.3&2.2 &   0.0&2.1 &   2.2&2.8 &   0.2&0.8\\
 1&a & --9.8&0.3 &   1.0&0.6 &   0.4&0.4 &   0.0&0.3 & --0.5&0.6 &   1.9&0.3 & --1.0&0.6 & --0.1&0.5 &   1.4&0.5 &   0.0&0.6 &   0.3&0.4\\
 2&  & --9.9&0.2 &   1.3&0.3 &   0.1&0.3 &   0.3&0.2 & --0.3&0.4 &   1.6&0.2 & --0.4&0.4 &   0.2&0.3 &   1.1&0.3 &   0.3&0.3 &   0.4&0.3\\
 2&a &--10.2&0.1 &   1.2&0.3 & --0.0&0.2 &   0.4&0.2 & --0.3&0.3 &   1.3&0.2 & --0.4&0.3 &   0.3&0.2 &   1.1&0.2 &   0.5&0.3 &   0.6&0.2\\
 3&  &--10.5&0.1 &   1.0&0.3 & --0.4&0.2 &   0.4&0.2 & --0.8&0.2 &   1.3&0.2 & --0.7&0.3 & --0.0&0.2 &   1.3&0.2 &   0.5&0.2 &   0.5&0.2\\
 3&a &--11.1&0.1 &   1.6&0.3 & --0.7&0.2 & --0.1&0.2 & --0.9&0.2 &   1.3&0.2 & --1.0&0.2 & --0.2&0.2 &   1.2&0.2 &   0.2&0.2 &   0.5&0.2\\
 4&  &--12.3&0.1 &   1.7&0.3 & --0.3&0.2 & --0.4&0.2 & --1.2&0.2 &   1.5&0.2 & --1.2&0.3 & --0.1&0.2 &   1.2&0.2 & --0.2&0.3 &   0.5&0.2\\
 4&a &--13.3&0.1 &   2.1&0.3 & --0.2&0.2 & --0.5&0.2 & --1.6&0.2 &   2.1&0.2 & --1.2&0.3 & --0.1&0.2 &   1.4&0.2 &   0.1&0.2 &   0.8&0.2\\
 5&  &--14.7&0.2 &   2.8&0.3 & --0.9&0.2 & --0.7&0.3 & --1.8&0.3 &   2.0&0.3 & --1.2&0.3 & --0.1&0.3 &   1.3&0.3 &   0.6&0.3 &   0.8&0.3\\
 5&a &--17.0&0.2 &   3.7&0.3 & --0.4&0.3 & --0.6&0.3 & --2.0&0.3 &   1.7&0.3 & --1.3&0.3 & --0.4&0.3 &   0.8&0.3 &   0.7&0.3 &   0.6&0.3\\
 6&  &--19.1&0.2 &   3.7&0.5 & --0.2&0.3 & --1.2&0.4 & --2.6&0.4 &   1.4&0.4 & --1.0&0.4 & --0.3&0.4 &   0.6&0.4 &   1.0&0.4 &   0.6&0.4\\
 6&a &--23.8&0.3 &   2.9&0.6 & --1.1&0.5 & --2.5&0.5 & --3.0&0.5 &   0.5&0.5 & --2.1&0.5 & --0.2&0.5 & --0.4&0.5 &   0.5&0.5 &   0.4&0.5\\
 7&  &--25.7&0.4 &   3.8&0.7 & --1.7&0.5 & --3.5&0.6 & --3.5&0.6 &   1.6&0.6 & --2.5&0.6 & --1.1&0.6 & --0.7&0.6 &   1.2&0.6 &   0.8&0.6\\
 7&a &--23.2&0.6 &   1.8&1.0 & --4.7&0.7 & --5.2&0.9 & --2.9&0.8 &   0.9&0.8 & --3.6&0.9 & --1.4&0.8 & --0.9&0.9 &   2.5&0.8 &   0.7&0.9\\
 8&  &--15.1&0.4 &   0.8&0.7 & --3.8&0.5 & --2.8&0.6 & --1.7&0.7 & --1.2&0.6 & --3.2&0.7 &   0.2&0.6 &   0.0&0.7 &   1.2&0.7 &   0.4&0.6\\
 8&a &--12.4&0.3 &   1.7&0.5 & --2.1&0.4 & --2.0&0.4 & --1.2&0.5 & --0.1&0.4 & --2.2&0.5 & --0.6&0.5 &   0.1&0.5 &   0.6&0.5 & --0.3&0.4\\
 9&  &--11.0&0.2 &   1.7&0.4 & --2.4&0.3 & --1.6&0.4 & --1.3&0.4 &   0.2&0.4 & --1.9&0.4 & --0.8&0.4 &   0.3&0.4 &   1.8&0.4 &   0.1&0.4\\
 9&a & --9.6&0.2 &   1.2&0.4 & --2.9&0.3 & --1.6&0.4 & --0.5&0.4 & --0.4&0.4 & --1.9&0.4 & --1.0&0.4 &   0.1&0.4 &   1.8&0.4 & --0.0&0.4\\
10&  & --8.4&0.2 &   1.2&0.4 & --2.8&0.3 & --1.1&0.3 &   0.8&0.4 & --0.5&0.3 & --3.0&0.4 & --0.7&0.4 &   0.0&0.3 &   0.8&0.4 &   0.0&0.3\\

  \enddata
\end{deluxetable}

The mode mixing caused by longitudinal variations of $\o{\pi}$ has
been corrected assuming the Fourier coefficients of $\o{\mu}_b$ are
due to this effect alone (cf.\ \Sec{tech-mix-low} and
\Sec{tech-num-mix}). The resulting mode-mixing corrected Fourier
coefficients are given in \Tab{results} and are discussed in the next
section.

\begin{deluxetable}{r@{}l@{\quad}
    r@{${\pm}$}l@{$\;\;$}r@{${\pm}$}l@{$\;\;$}r@{${\pm}$}l  
    r@{${\pm}$}l@{$\;\;$}r@{${\pm}$}l@{$\;\;$}r@{${\pm}$}l  
    r@{${\pm}$}l@{$\;\;$}r@{${\pm}$}l                       
    r@{${\pm}$}l@{$\;\;$}r@{${\pm}$}l}                      
  \ifpreprint
     \tabletypesize{\footnotesize}
  \else
     \tabletypesize{\scriptsize}
  \fi
  \tablewidth{0pt}
  \tablecaption{Results after Correcting for Mode Mixing
    \label{tab:results}}
  \tablehead{
    \multicolumn{2}{c}{bin}     &
    \multicolumn{2}{c}{$\o{\mu}_{U0}$} &
    \multicolumn{2}{c}{$\o{\mu}_{V0}$} &
    \multicolumn{2}{c}{$\o{\mu}_{W0}$} &
    \multicolumn{2}{c}{$\tilde{A}$}   &
    \multicolumn{2}{c}{$\tilde{B}$}   &
    \multicolumn{2}{c}{$\tilde{C}$}   &
    \multicolumn{2}{c}{$c_3$}   & \multicolumn{2}{c}{$s_3$}   &
    \multicolumn{2}{c}{$c_4$}   & \multicolumn{2}{c}{$s_4$} 
    }
  \tablecolumns{22}
  \startdata
 1&  &  11.7&2.9 &  15.3&2.1 &  11.9&1.7 &  10.9&2.2 &--11.6&1.5 & --0.1&2.5 &   0.3&2.0 & --3.6&2.7 & --2.7&2.4 &   1.0&2.3\\
 1&a &  13.4&1.7 &  15.5&1.3 &  11.3&1.0 &   8.4&1.4 &--12.7&1.0 & --2.0&1.5 &   1.6&1.2 & --4.5&1.8 & --1.1&1.3 &   4.8&1.6\\
 2&  &  13.1&0.7 &  15.9&0.8 &  10.0&0.3 &   9.6&0.8 &--13.3&0.6 & --2.1&1.0 &   1.4&0.8 & --5.1&1.0 & --0.8&0.8 &   2.7&1.1\\
 2&a &  14.4&0.6 &  18.1&0.6 &  10.3&0.2 &  11.6&0.7 &--12.9&0.5 & --2.4&0.8 &   1.7&0.7 & --3.9&0.8 & --0.6&0.7 &   0.9&0.8\\
 3&  &  15.0&0.7 &  18.3&1.0 &  11.0&0.4 &  12.0&1.0 &--13.8&0.8 & --4.0&1.4 &   2.6&0.7 & --5.3&1.6 & --0.0&1.0 &   2.2&1.5\\
 3&a &  15.7&0.8 &  19.0&0.9 &  11.2&0.4 &  12.7&1.0 &--14.4&0.8 & --4.1&1.2 &   2.4&0.9 & --5.6&1.4 &   1.2&0.9 &   0.9&1.4\\
 4&  &  17.8&0.9 &  22.7&0.8 &  11.9&0.3 &  11.3&1.1 &--13.9&0.7 & --2.4&1.0 &   2.2&1.0 & --3.0&1.1 &   0.3&1.0 &   0.2&1.1\\
 4&a &  19.1&1.0 &  26.8&1.0 &  13.4&0.3 &   9.0&1.3 &--14.4&0.8 & --2.0&1.1 &   2.2&1.2 & --3.7&1.2 & --0.9&1.1 &   1.3&1.3\\
 5&  &  21.8&1.0 &  29.7&1.1 &  15.2&0.4 &   9.8&1.4 &--16.0&0.9 & --3.2&1.1 &   1.5&1.3 & --4.4&1.2 &   0.7&1.2 &   0.8&1.3\\
 5&a &  26.7&1.1 &  36.5&1.1 &  18.2&0.4 &  12.4&1.4 &--15.9&0.9 & --4.2&1.2 &   2.0&1.3 & --3.8&1.3 & --1.3&1.3 & --0.4&1.3\\
 6&  &  30.2&1.3 &  43.1&1.3 &  21.1&0.5 &  13.8&1.8 &--14.2&1.2 & --4.4&1.5 &   3.0&1.6 & --3.2&1.6 & --1.5&1.6 &   0.7&1.7\\
 6&a &  37.0&1.8 &  61.4&1.6 &  25.9&0.6 &  14.8&2.3 &--12.8&1.5 & --8.6&2.0 &   2.5&2.1 & --0.2&2.1 &   2.3&2.0 &   4.7&2.3\\
 7&  &  40.3&2.1 &  72.2&1.9 &  28.2&0.6 &  14.5&2.7 &--12.9&1.8 & --7.8&2.3 &   4.5&2.5 &   0.9&2.4 &   2.9&2.5 &   1.7&2.5\\
 7&a &  39.4&3.1 &  73.2&2.6 &  28.4&1.0 &  11.9&4.2 &--14.4&2.7 & --2.7&3.3 &   6.8&3.7 &   2.7&3.5 &   2.9&3.9 & --5.4&3.7\\
 8&  &  29.0&3.0 &  53.9&2.5 &  21.2&0.9 &  12.7&3.7 &--16.4&2.5 & --4.4&3.3 &   3.2&3.3 &   2.3&3.6 &   0.2&3.6 & --2.1&3.6\\
 8&a &  22.8&2.5 &  41.4&2.1 &  15.4&0.7 &  12.7&3.1 &--17.7&2.1 & --5.5&3.0 &   5.7&2.6 & --2.5&3.4 &   0.1&3.1 & --0.0&3.1\\
 9&  &  19.3&2.4 &  36.7&1.7 &  12.5&0.6 &  14.2&2.2 &--17.4&1.7 &--10.2&2.5 &   6.6&2.4 & --1.6&2.6 &   0.9&2.5 &   0.6&2.4\\
 9&a &  17.6&2.8 &  35.3&1.9 &  11.7&0.6 &  14.2&2.3 &--17.3&1.8 & --9.8&3.0 &   5.4&3.0 &   1.6&3.1 &   1.8&3.1 &   2.0&2.7\\
10&  &  13.4&2.0 &  28.5&1.8 &   9.5&0.6 &  13.8&2.1 &--16.2&1.5 & --9.0&2.3 &   3.2&2.3 &   0.2&2.7 &   1.8&2.5 &   5.6&2.4\\

  \cutinhead{with $\kappa$-$\sigma$-clipping}
 1&  &  15.1&3.0 &  16.7&1.7 &  11.0&1.6 &   9.7&1.4 &--13.3&1.2 & --1.5&2.0 &   2.4&1.7 & --4.1&2.0 &   1.1&1.6 &   3.8&1.7\\
 1&a &  13.4&0.8 &  15.6&0.8 &   9.9&0.3 &   9.8&0.8 &--13.1&0.7 & --1.4&1.1 &   1.9&0.8 & --4.8&1.1 &   0.4&0.9 &   2.8&0.9\\
 2&  &  13.6&0.6 &  16.3&0.6 &   9.9&0.2 &  10.6&0.6 &--13.0&0.5 & --1.8&0.8 &   1.8&0.6 & --4.4&0.8 & --0.8&0.7 &   1.8&0.7\\
 2&a &  13.9&0.5 &  17.7&0.5 &  10.2&0.1 &  11.2&0.5 &--13.1&0.4 & --2.1&0.7 &   1.5&0.5 & --4.0&0.7 & --0.5&0.6 &   1.3&0.6\\
 3&  &  14.2&0.5 &  19.3&0.5 &  10.5&0.1 &  12.0&0.5 &--13.3&0.4 & --2.4&0.6 &   1.4&0.5 & --3.5&0.6 & --0.5&0.6 &   0.9&0.6\\
 3&a &  15.0&0.5 &  20.1&0.4 &  11.2&0.1 &  11.9&0.5 &--13.9&0.4 & --3.6&0.6 &   0.9&0.5 & --3.3&0.6 & --0.0&0.6 &   0.7&0.6\\
 4&  &  17.5&0.5 &  21.4&0.5 &  12.3&0.1 &  11.0&0.6 &--13.6&0.4 & --4.0&0.6 &   1.1&0.6 & --2.7&0.6 &   0.3&0.6 &   1.5&0.6\\
 4&a &  20.0&0.5 &  23.8&0.5 &  13.3&0.1 &  10.9&0.6 &--13.7&0.4 & --3.9&0.6 &   2.4&0.6 & --2.8&0.6 &   0.6&0.6 &   1.9&0.6\\
 5&  &  22.5&0.6 &  26.3&0.6 &  14.8&0.2 &  11.5&0.7 &--14.3&0.4 & --5.3&0.6 &   2.2&0.7 & --2.6&0.7 &   0.0&0.7 &   1.2&0.7\\
 5&a &  26.2&0.7 &  32.9&0.7 &  17.1&0.2 &  12.0&0.8 &--14.8&0.5 & --6.2&0.8 &   1.5&0.8 & --1.4&0.8 & --0.6&0.8 &   1.3&0.8\\
 6&  &  28.8&0.9 &  39.4&0.9 &  19.2&0.2 &  12.5&1.1 &--14.8&0.7 & --5.6&1.0 &   2.2&1.1 & --0.7&1.0 & --0.8&1.0 &   2.1&1.0\\
 6&a &  33.5&1.2 &  53.1&1.1 &  23.8&0.3 &  13.0&1.5 &--14.0&1.0 & --8.6&1.3 &   1.2&1.4 &   0.0&1.4 &   1.9&1.4 &   2.5&1.4\\
 7&  &  37.2&1.4 &  61.8&1.3 &  25.7&0.4 &  10.9&1.7 &--15.5&1.1 & --8.3&1.5 &   4.1&1.6 &   2.0&1.6 &   0.7&1.6 &   1.4&1.7\\
 7&a &  34.9&2.1 &  61.5&1.8 &  23.2&0.6 &  10.3&2.6 &--16.2&1.7 & --4.4&2.2 &   6.4&2.4 &   3.9&2.4 & --0.7&2.4 & --1.6&2.5\\
 8&  &  24.2&1.6 &  39.9&1.4 &  15.1&0.4 &  16.7&1.8 &--13.8&1.2 & --5.5&1.8 &   4.4&1.7 &   3.6&2.1 &   0.7&1.8 & --2.1&1.9\\
 8&a &  19.6&1.2 &  32.5&1.0 &  12.5&0.3 &  14.2&1.3 &--15.4&0.9 & --7.0&1.3 &   4.9&1.3 &   1.5&1.5 &   0.6&1.3 &   1.5&1.4\\
 9&  &  17.5&1.1 &  31.7&0.9 &  11.0&0.2 &  12.8&1.2 &--17.3&0.8 & --9.6&1.2 &   4.3&1.2 &   1.4&1.3 & --0.3&1.2 &   1.7&1.2\\
 9&a &  16.1&1.1 &  29.7&0.9 &   9.6&0.2 &  13.8&1.3 &--17.1&0.9 & --9.8&1.2 &   4.6&1.2 &   1.6&1.3 & --0.2&1.3 &   1.3&1.3\\
10&  &  12.6&0.9 &  25.2&0.8 &   8.4&0.2 &  12.9&1.1 &--16.6&0.8 & --9.0&1.1 &   3.1&1.1 &   0.5&1.2 &   0.6&1.1 &   3.8&1.1\\

  \enddata
\end{deluxetable}

\subsubsection{Secular Parallax and Asymmetric Drift}
One might estimate the mean parallax $\o{\pi}_0$ and the azimuthal
asymmetric drift $v_{a\varphi}$ from the reflex of the solar motion
and the assumption that there are no radial and vertical components to
the asymmetric drift, i.e.\ 
\begin{equation} \label{eq:solar-motion}
  \B{v}_\odot = (U_0,\,V_0+v_{a\varphi},\,W_0).
\end{equation}
Here $(U_0,V_0,W_0)$ is the solar motion with respect to the local
standard of rest (LSR), which we take from DB98 to be
$(10.00\pm0.36,\,5.25\pm0.62,\,7.17\pm0.38)\,\kms$. Inserting
\eq{solar-motion} into \eqn{pm-sun-zero}, we can solve for $\o{\pi}_0$
and $v_{a\varphi}$ in two different ways yielding
\begin{mathletters} \label{eq:pi-sec}
  \begin{eqnarray}
    \label{eq:pi-u}
    \o{\pi}_0 &=& \o{\pi}_U 
    \equiv \frac{\o{\mu}_{U0}}{ U_0},
    \qquad
    v_{a\varphi} = U_0 \frac{\o{\mu}_{V0}}{\o{\mu}_{U0}}-V_0;          \\
    \label{eq:pi-w}
    \o{\pi}_0 &=& \o{\pi}_W \equiv \frac{\o{\mu}_{W0}}{ W_0},
    \quad\;\;
    v_{a\varphi} = W_0  \frac{\o{\mu}_{V0}}{\o{\mu}_{W0}}-V_0.
  \end{eqnarray}
\end{mathletters}

\subsubsection{Intrinsic Colors?} \label{sec:Intrinsic_Colors}
We have mentioned before that extinction is partly to blame for the
non-uniformity in the distribution of stars.  In addition,
interstellar dust reddens the intrinsic colors of stars. To estimate
these intrinsic colors, we assume that all stars of a given color in
our sample have the distance equal to the mean parallax \eq{pi-w}.
With the standard extinction law ($R_V=3.1$) the intrinsic color can
be approximated as:
\begin{eqnarray}
  (\bv)_0 &=& (\bv)_{\mathrm{obs}}-E_{B-V} \\ \nonumber
  &\approx& (\bv)_{\mathrm{obs}}-\frac{1\times d}{R_V},
  \nonumber
\end{eqnarray}
where we use the subscript 0 for colors corrected assuming an
average extinction of 1 mag per kpc \citep[e.g.,][]{Cea98}. In
\Tab{bins}, we include these estimates of the intrinsic colors, as
well as an estimate for the average distance ($1/\o{\pi_W}$).

Red clump stars with intrinsic $\bv\in[0.9,1.1]$ have absolute
magnitudes similar to A-type main-sequence stars with intrinsic
$\bv\in [-0.1,0.1]$ (e.g., DB98), so that the average distance to the
A-type and red-clump stars should be similar.  Inspecting \Tab{bins},
it is obvious that the no-extinction case does not conform to this
this expectation at all. On the other hand, when using the
extinction-corrected colors, $(\bv)_0$, to determine the observed
color ranges for the two sub populations, we find almost identical
distances for the red clump and A type stars. We thus conclude that,
in the average, the extinction corrected colors are close to the
intrinsic colors of our target stars. For the remainder of this paper
we assume that $(\bv)_0$ is a good approximation for the intrinsic
color. But note that our conclusions do not critically depend on this
assumption.
\section{Discussion of the Results}     \label{sec:disc}
In this section, we discuss only the results obtained with
$\kappa$-$\sigma$-clipping. These have smaller errors than the results
obtained using all stars and there are only minor systematic
deviations between the two sets.

\subsection{Initial Considerations}
Before trying to understand and interpret the results derived from the
proper motions, we must be aware of the kind of stars we are dealing
with in the various color bins.  As discussed in \S\Sec{adrift} and
\ref{sec:sys-r}, it is well known that the kinematics of stars changes
systematically with age: the age-velocity relation (AVR).  A critical
age for a stellar population is 1.5-2~Gyr.  The kinematics of youngest
stars still carry a significant imprint of the initial conditions,
while older stars have had time to reach an equilibrium with the
large-scale potential of the Milky Way \cite[e.g.,][]{M74,GM77}.
Since we are interested primarily in the large-scale properties of the
Galactic potential, it is important to be able to estimate the ages of
the stars in our samples.  The ACT data-base does not allow for
sophisticated age estimates of individual stars, but we can estimate
the fraction of ``young'' stars in each of the color bins we use.
Stars bluer than $(\bv)_0 = 0.25$ (bins 1--5a) have main-sequence
lifetimes smaller than 1.5~Gyr, and we expect these stars to exhibit
kinematics appropriate for young stellar populations.

\subsubsection{Sample Properties} \label{sec:Sample_properties} 
One can estimate the fraction of stars younger than 1.5~Gyr if we
assume a rapid post-main-sequence evolution and a constant
star-formation rate (SFR) over the last several Gyr.  For stars with a
main-sequence lifetime of 8.4~Gyr [bin 7, $(\bv)_0 \sim0.6$],
$\sim$18\% of the stars will be younger than 1.5~Gyr.

Keeping in mind that the fraction of young stars decreases along the
main sequence, we might interpret the gradual decrease in
non-uniformity of the stellar number density (\Fig{nl}) as a
decreasing fraction of young stars.  In \Fig{nonuni} we present a
measure, $\delta n(\ell)$, of the non-uniformity of the number-density
distribution. In fact, we can use $\delta n(\ell)$ as a proxy for the
fraction of young stars.

If we approximate that $\delta n(\ell)$ has just two distinct values
$\delta n(\ell)_Y$ and $\delta n(\ell)_O$ for `young'
($\tau<1.5\,$Gyr) and for `old' stars, we can estimate the fraction
of young stars as
\begin{equation} \label{eq:F_Y_n_ell}
  f_Y \approx \frac{\delta n(\ell)   - \delta n(\ell)_O}
                   {\delta n(\ell)_Y - \delta n(\ell)_O}.
\end{equation}
For the young stars, we can take the weighted average of the bins with
$(\bv)_0\la 0.25$, yielding $\delta n(\ell)_Y=0.126$.  For the old
stars, we can use $f_Y=0.18$ for bin \#7 and invert \eqn{F_Y_n_ell} to
obtain $\delta n(\ell)_O\sim0.024$.  We apply \eqn{F_Y_n_ell} to
estimate the fraction of young stars among the giants. The results are
tabulated in \Tab{bins} for the case of 1~magnitude extinction per
kpc. (If no extinction correction is used, the $f_Y$ values decrease
by $\sim$25\% over the values listed.)

Thus, at the blue end, the stars are both young and bright, with ages
up to to a few rotation periods of the Galaxy, and distances out to
2\,\kpc.  For ever redder stars up to $\bv\simeq0.7$, there is a
gradual change to fainter, hence nearer, and on average, older stars
(cf.  \Tab{bins}).  As a consequence, the internal kinematics and the
averaging volume of the stars change; the first due to the
kinematics-age dependence, the latter simply because of the
distance-color relation on the main-sequence (\Tab{bins}).

Beyond $\bv\simeq0.7$, there is a more abrupt change of stellar
properties of the sample, because non-main-sequence stars take over to
dominate the sample.  These stars are both brighter and, on average,
younger than the dwarfs, changing again both the kinematics and the
averaging volume.  If we take the non-uniformity as an age indicator,
\Fig{nonuni} clearly shows that the stars with $\bv\approx1$ are younger
than both the bluer dwarfs and the redder giants. In fact, the $\delta
n(\ell)$ value at $\bv=1$ equals that of A-type stars, suggesting an
average age of only a few hundred million years for stars in this
color range.

Using the $(\bv)_0$ colors listed in \Tab{bins} and a color magnitude
diagram representative of the Solar neighborhood (e.g., DB98, their
Fig.~1), we identify the stars in color bins 7a to be sub-giants.  Bin
8 is a mixture of sub-giant, giant and red-clump stars.  Red-clump
stars dominate bins 8a, 9, and 9a.  Only the last color bin, \#~10,
predominantly comprise red-giant stars.

Our $\delta n(\ell)$ analysis above indicates a significant fraction
of young stars in all red color bins (see \Tab{bins}).  This can be
understood in the context of ongoing star-formation activity in the
Solar neighborhood and the theory of stellar post-main-sequence
evolution \citep{sei87,col98,gir98}.

It is worth mentioning that at $(\bv)_0>1$, the difference in luminosity
between dwarfs and (sub) giant stars is larger than about 3\,mag such that the
number of red dwarfs beyond that color is negligible, in particular after
$\kappa$-$\sigma$-clipping has been applied.

\subsubsection{Expectations for the Kinematics}
From our considerations in \Secsto{adrift}{sys-r}, we expect the
changes in kinematics and averaging volume to be reflected in the
the Oort constants {\em measurable\/} for the stars.  Thus, already
without the unpleasant effect of mode mixing, we expect the Oort
constants to change gradually blue-ward of $\bv\simeq0.7$ and red-ward
of $\bv\simeq1$, while there might be a more abrupt change between
these two color ranges.

An important question will then be: which stars give us the ``true
Oort constants''?  The young blue stars ($\tau_{MS}\la 1.5$ Gyr) are
not likely to be in equilibrium as they still exhibit kinematics
associated with their birth places.  Thus, their streaming velocity
field likely deviates systematically from that created by closed
orbits, and their distances and velocities are correlated.  Both
effects cause systematic errors when interpreting the $m=0$ and 2
coefficients as the Oort constants. The dwarfs at intermediate colors
($0.25\la(\bv)_0\la0.7$) probe a very local volume, the secular
parallax is estimated to be less than 500~pc, and it is likely that
their measurable streaming field is affected by local anomalies
(\Secsto{sys-depth}{sys-r}).

The red-clump comprises a mixture of stars of various ages, which
judged from their longitudinal distribution is quite affected by
spiral arms and thus subject to the same objections as the blue stars.
Only samples of giants redder than $(\bv)_0\approx1.2$ (our color bin \#10)
might be both old enough and distant enough to be unaffected by
non-equilibrium effects or local anomalies.

\subsection{The Fourier Coefficients Measured} 
We plot in \Fig{Oort.vs.BV} as function of intrinsic color $(\bv)_0$
the results for the Oort constants, the inverse mean parallax, the
asymmetric drift, $\tilde{A}-\tilde{B}$, and the $m=3,4$ terms of
$\o{\mu}_{\ell^\star}$ before (open circles) and after (solid squares)
mode-mixing correction.

\begin{figure*}[t]
   \ifpreprint
      \centerline{
         \resizebox{85mm}{!}{\includegraphics{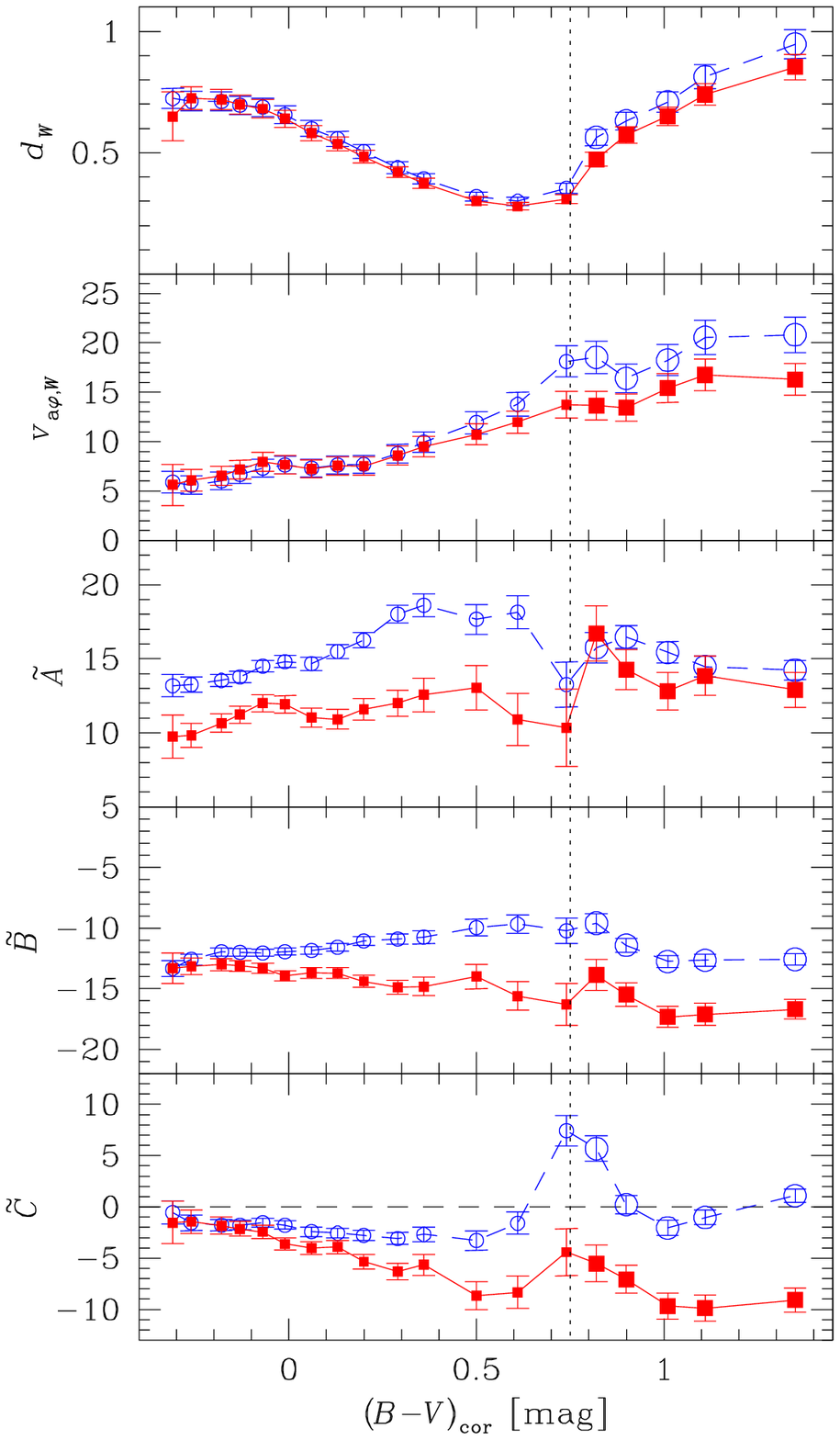}}
         \hspace{1em}
         \resizebox{85mm}{!}{\includegraphics{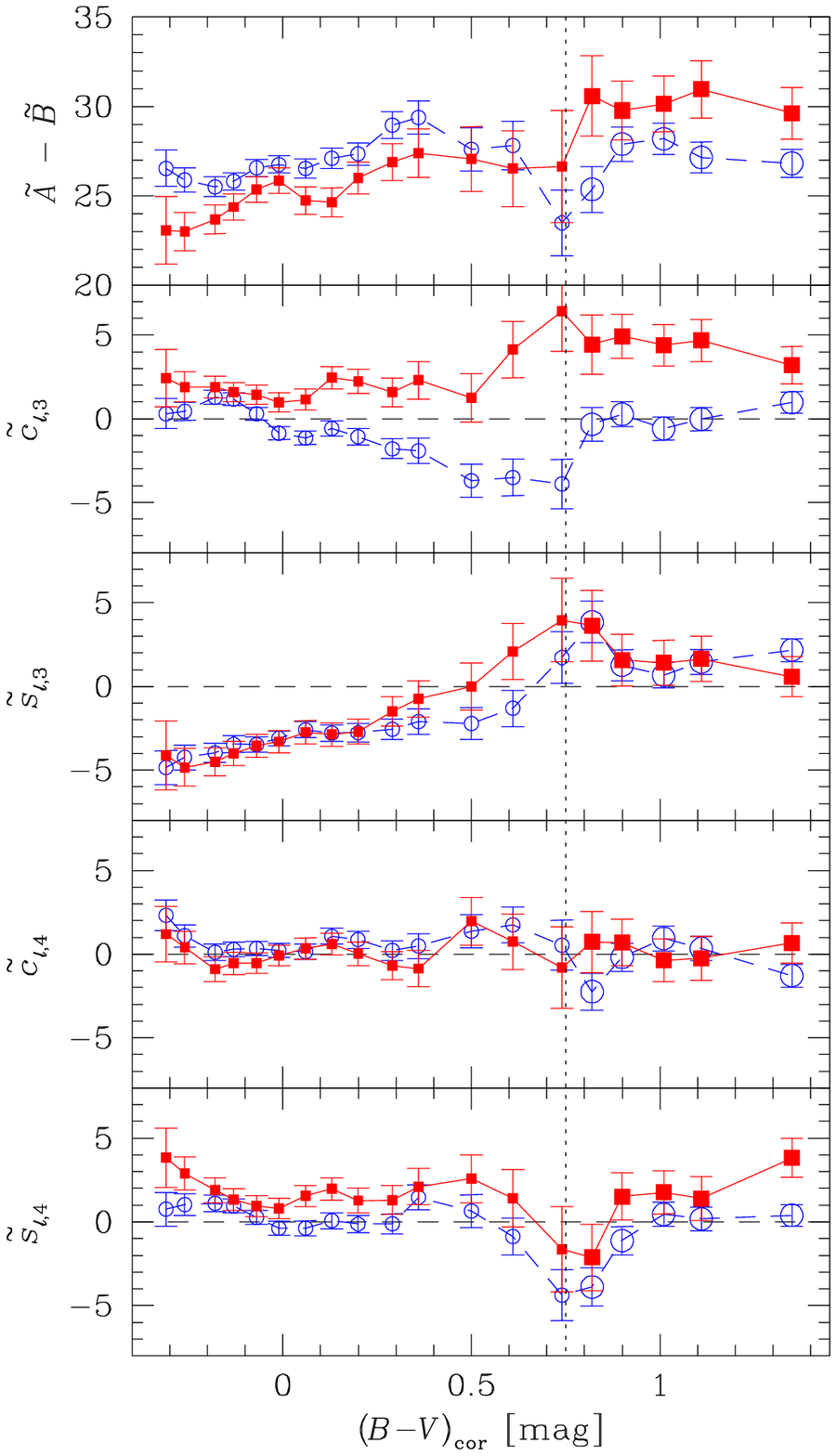}}
      }
   \else
      \centerline{
         \resizebox{85mm}{!}{\plotone{f5a}}
         \hspace{1em}
         \resizebox{85mm}{!}{\plotone{f5b}}
      }
   \fi
   \input{ms_fc05}
\end{figure*}
Obviously, there is a significant difference between the raw and
mode-mixing corrected values.  From our discussion in
\Sec{Mode_Mixing_Correct_Q}, we expect the ``true'' values will be
close to the mode-mixing corrected values, but we cannot entirely
exclude the raw values as a possibility.  We will consider the
difference between the raw and mode-mixing corrected values as an
upper limit to the systematic error involved.  Regardless of this
difficulty, several inferences can be made.

First, there is an obvious discontinuity in kinematic properties at
$(\bv)_0\approx0.75$. This discontinuity is reflected in all Fourier
coefficients of $\o{\mu}_{\ell^\star}$ raw or mode-mixing corrected.
This is unlikely to be caused predominantly by a difference in
distance, since there are no significant correlations between the
Fourier coefficients and $\o{\pi}^{-1}$ (not shown).  Presumably more
important is that these stars span the range in ages where the
gradient in the age-velocity relation is large.  The unstable nature
of the Fourier coefficients in the red-clump region illustrates that
it is very difficult to determine the true value of the Oort constants
if the ages of the tracer stars are ill-determined.
\begin{figure}[th]
   \ifpreprint
      \centerline{\resizebox{85mm}{!}{\includegraphics{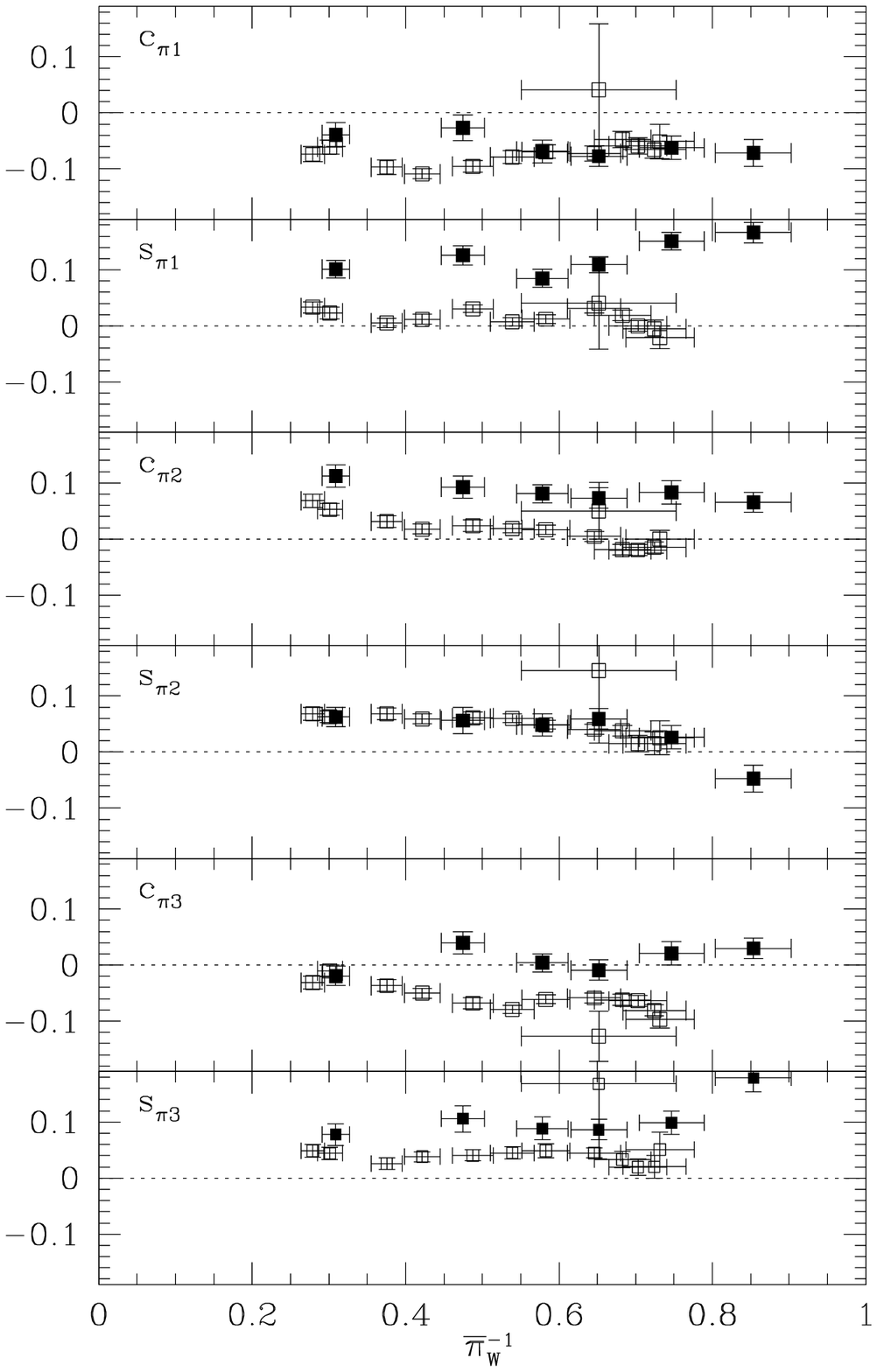}}}
   \else
      \centerline{\resizebox{85mm}{!}{\plotone{f6}}}
   \fi
   \input{ms_fc06}
\end{figure}
\begin{figure}[t] 
   \ifpreprint
      \centerline{\resizebox{80mm}{!}{\includegraphics{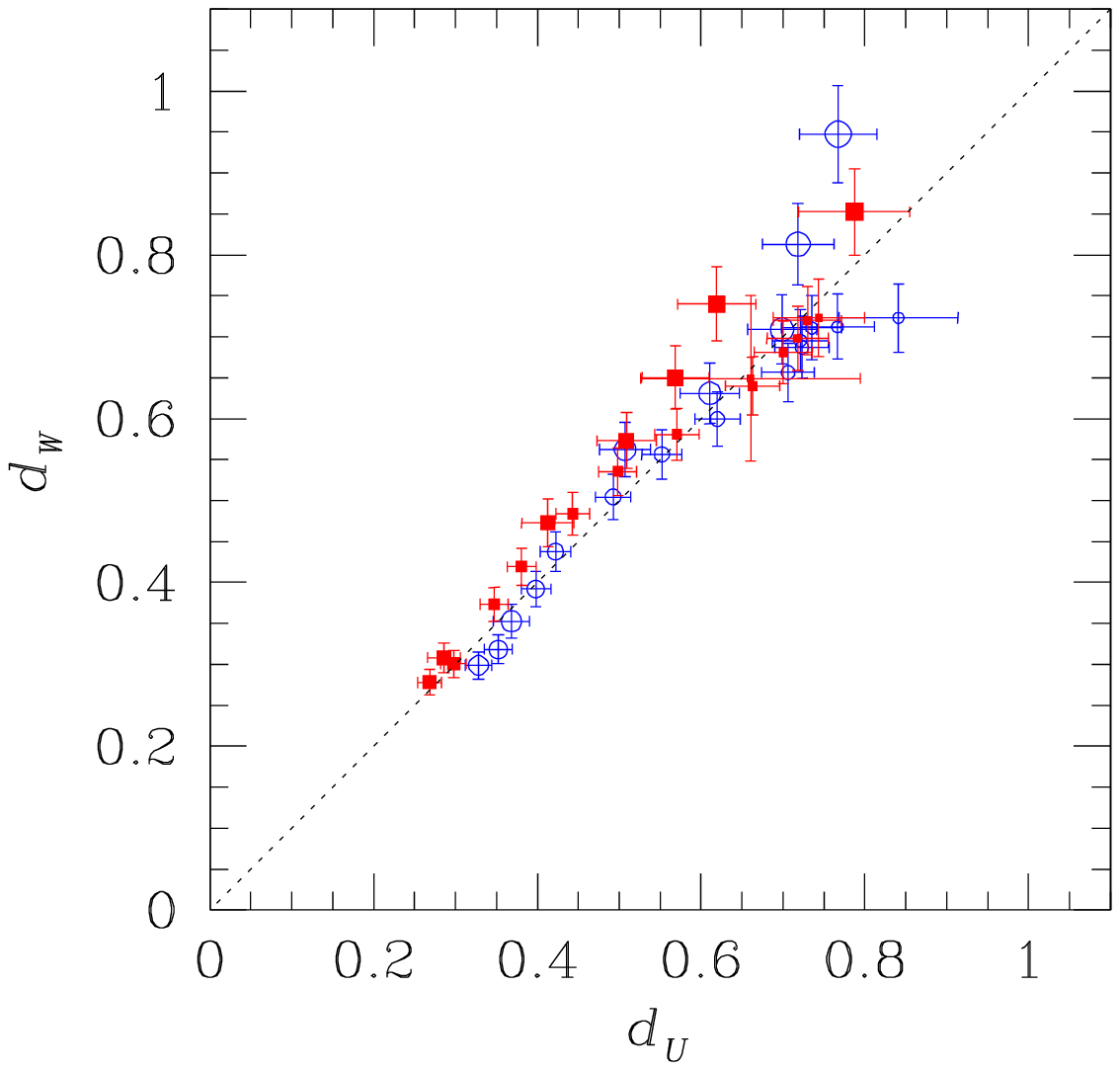}}}
   \else
      \centerline{\resizebox{85mm}{!}{\plotone{f7}}}
   \fi
   \input{ms_fc07}
\end{figure}
Second, the asymmetric drift and $\tilde{A}-\tilde{B}$ are only weakly
affected by the systematic errors. $\tilde{A}-\tilde{B}\approx26$ for
blue stars (uncorrected for mode-mixing), which is in nice agreement
with $\tilde{A}-\tilde{B}=27$ measured from Cepheids by \citet{fw97}.
There is a trend towards larger values for redder stars.  At the red
end, we again notice the strong discontinuity in the red-clump region
$(\bv)_0=0.75\pm0.15$. The red giant bin (\#10) has
$\tilde{A}-\tilde{B}=26.7\pm0.7$ and $29.5\pm1.4$ for the raw and
mode-mixing corrected values, respectively. These values are
consistent with a Galactic circular frequency of $28\pm2$ as derived
from the proper motion for Sgr~A$^\star$ \citep{Rea99,BS99}.

Third, the values for $\tilde{A}$ before and after mode-mixing
correction differ by just $1\sigma$ for the red giant stars:
$\tilde{A}$ is between $14.2\pm0.6$ and $12.9\pm1.1$.

Fourth, except for the very blue stars, there is a large systematic
uncertainty for $\tilde{B}$.  The red giant bin has $\tilde{B}$ values
between --16.6 and --12.5.

Fifth, there are several clear evidences for deviations from
axisymmetric equilibrium: non-zero $\tilde{C}$ and $\sin3\ell$ terms
(the latter only for blue stars). Note, in particular, that after
mode-mixing correction $\tilde{C}$ has the same sign for all stars.

Finally, while the asymmetric drift is increasing with color
independent of whether it has been derived from the raw or mode-mixing
corrected coefficients, the curves derived from the $U$ and the $W$
motion differ clearly (not shown).  In the latter case, the increase
is more gentle and reaches only $v_{\rm a\varphi} \approx 16\,\kms$, whereas
the $U$-derived coefficients yield $v_{\rm a\varphi}$ 2-4\,\kms larger.
Note that we do not expect an exact correspondence with the asymmetric
drift derived by DB98 because our samples comprise mixes of
sub-populations that differ from the local Hipparcos sample, in
particular at the red end.

\subsection{Mode Mixing}
\placefigure{fig:mix}
\subsubsection{Evidence for Mode Mixing}
In \Fig{mix}, we plot the Fourier coefficients $(c_\pi,s_\pi)_m$,
derived from the vertical proper motions utilizing \eqn{mix-mub-low},
versus the inverse mean parallax estimated from the average vertical
proper motion via \eqn{pi-w}. Clearly, these coefficients deviate
significantly from zero proving that mode mixing is present and
non-negligible, i.e.\ that the Oort constants obtained without
correction are systematically in error. There is a clear dichotomy
between main-sequence stars ($\bv<0.8$; open symbols) and
non-main-sequence stars ($\bv>0.8$; closed symbols): most mode-mixing
coefficients are similar within each color group but differ between
them. This indicates different spatial distributions, necessitating
separate analyses.


\subsubsection{Is our Mode-Mixing Correction Correct?} 
\label{sec:Mode_Mixing_Correct_Q}
We have tried to correct for the mode mixing by employing the vertical
proper motion and the technique described in \Sec{tech-mix-low} and
\Sec{tech-num-mix}.  However, as discussed in \Sec{mix.dist}, there
are possible caveats in this method, in particular neglecting
contributions from higher-order terms (cf.\ 
\Secsto{sys-depth}{sys-r}). One possible check on the consistency of
the results obtained is a comparison of the secular parallaxes
estimated from the solar radial and vertical motions via \eqs{pi-sec}.
\Fig{pi.vs.pi} compares the estimate $\o{\pi}_U^{-1}$, which is
affected by mode mixing, with $\o{\pi}_W^{-1}$, which is not, before
(circles) and after (squares) mode-mixing correction has been applied.
Obviously, in both cases a systematic deviation of 5--10\% exists
between the two estimates, but with opposite signs. An exception are
the very early-type stars ($\bv\la0.2$), which are known to deviate
from equilibrium (DB98).

Note that while the deviations from the dotted line increase with
distance, they do so less for the mode-mixing corrected results.  This
indicates that the neglect of higher-order terms, which should
introduce systematic errors at large distances, cannot have introduced
significant errors.

There are two possible explanations for those differences.  First, the
ratio $U_0/W_0$ may be different for this sample than for the more
local sample of Hipparcos stars utilized by DB98. In this case, we
cannot make any statement, whether our mode-mixing correction works or
not.

Second, if $U_0/W_0$ is equal the DB98 value, we can determine an
alternative mode-mixing solution in which we force equality between
$\o{\pi}_U$ and $\o{\pi}_W$. In this case, the radial and tangential
proper motions are given by
\begin{eqnarray}
  \o{\mu}_{U0,W} &=& U_0 \o{\pi}_0 = \frac{U_0}{W_0} \times \o{\mu}_{W0}
  \nonumber \\
  \o{\mu}_{V0,W} &=& \frac{s_{\pi 2} \times \o{\mu}_{U0,W} - c_{l1}}
  {1 + c_{\pi 2}}
  \nonumber
\end{eqnarray}
from \eqs{pi-sec} and \eq{mix-mul}, respectively (we have dropped the
latitude dependence). When using these relation for $\o{\mu}_{U0}$ and
$\o{\mu}_{V0}$ and solve the mode-mixing \eqs{mix-mul} directly, we
find that the so-determined Oort constants do not differ substantially
from our previous mode-mixing corrected values. This indicates that
the slight difference between $d_U$ and $d_W$ does not signify a
substantial problem for the mode-mixing scenario.

Since there is overwhelming evidence for the reality of the
mode-mixing effect (\Fig{mix}) and the details of the mode-mixing
procedure appear to be irrelevant, we conclude that the mode-mixing
solutions are robust. In particular, we are confident (and hope to
have convinced the reader, too) that the remaining systematic errors
of the corrected results are signigicantly smaller than for the raw
values. In our subsequent analysis, we will concentrate on the
mode-mixing corrected values, but will also show the raw results for
comparison.
\subsection{Can we make sense of these results?}
\label{sec:Making_sense}
In \Sec{prac}, we discussed various potential origins for deviations
of the measured Oort constants from their ``true'' values. These
deviations, which can be several \kmskpc, originate from departure of
the Milky Way from a smooth axisymmetric equilibrium, and may depend
on velocity dispersion and mean depth of the stellar population
considered.  Variations of this order are seen in our results in
\Fig{Oort.vs.BV}. While a detailled interpretation of the measured
variations of the Oort constants is beyond the scope of this paper, we
may nontheless examine
whether we can single out a dominant cause%
\footnote{We like to mention at this point that any systematic error
  in the cataloged proper motions also adds to these variations.
  However, the variations persist when performing the same analysis on
  the Tycho Reference Catalogue \citep{TRC} or the Tycho-2 Catalog
  \citep{Tyc2b}. This latter catalog is based on the same data as the
  ACT, but its proper motions are derived in a different manner, so
  that it seems unlikely that the proper motions are systematically in
  error \citep[see also][]{UWM2000}.}.
\begin{figure*}[t]
   \ifpreprint
      \centerline{
         \resizebox{85mm}{!}{\includegraphics{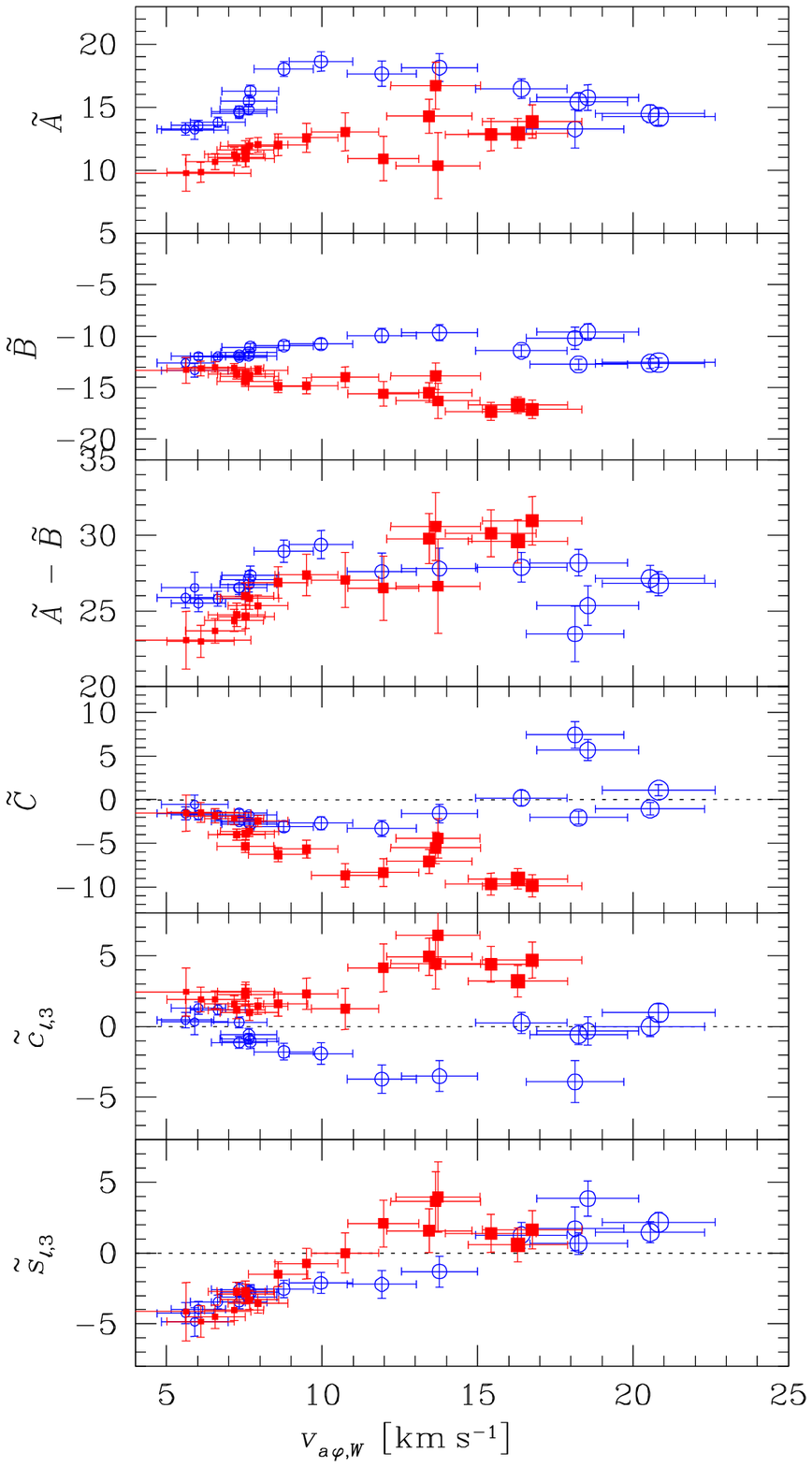}}
         \hspace{1em}
         \resizebox{85mm}{!}{\includegraphics{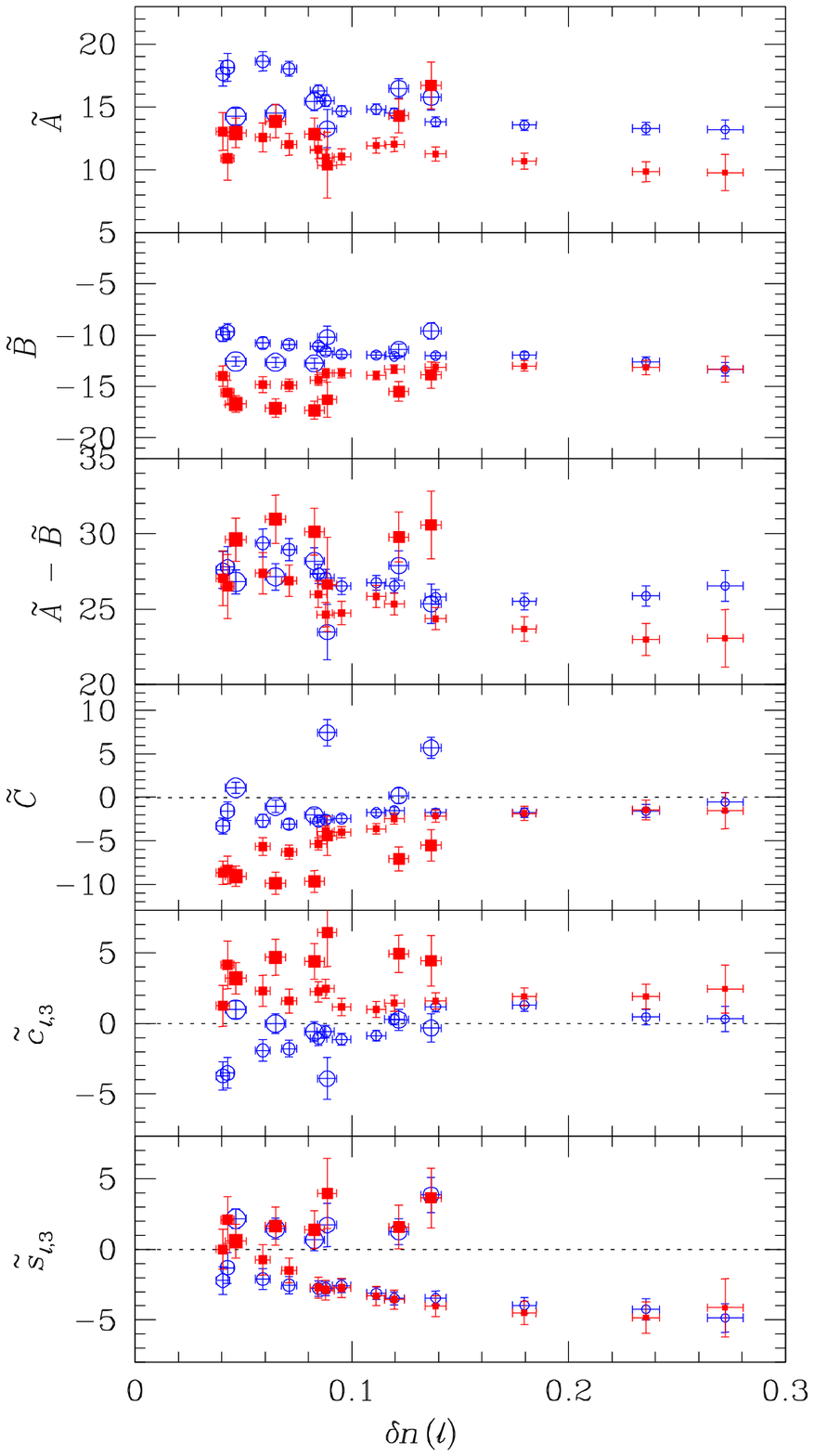}}
      }
   \else
      \centerline{
         \resizebox{85mm}{!}{\plotone{f8a}}
         \hspace{1em}
         \resizebox{85mm}{!}{\plotone{f8b}}
      }
   \fi
   \input{ms_fc08}
\end{figure*}

To disentangle the factors that may contribute to the variation of
$\tilde{B}$ with $v_{a\varphi}$, we first compare three
sub-populations in \Tab{sub_pops}. The bluest and reddest stars have
similar distances but very different ages, and hence also different
velocity dispersions.  As the third, we take the reddest main-sequence
stars at $(\bv)_0\sim 0.6$, which are much nearer.


\begin{table}[t]
  \begin{center}
    \scshape
    \refstepcounter{table} \label{tab:sub_pops}
    Table \arabic{table} \\
    Properties of some critical sub-populations
  \end{center}
  \begin{tabular*}{\columnwidth}{c@{\extracolsep{\fill}}
                                 c@{\extracolsep{\fill}}
                                 c@{\extracolsep{\fill}}
                                 r@{\extracolsep{\fill}}
                                 r@{\extracolsep{\fill}}
                                 l@{\extracolsep{\fill}}
                                 c@{\extracolsep{\fill}}
                                 c}
    \hline\hline
    \\[-1.5ex]
     $(\bv)_0$ & $d$ & $\sigma_*$ &
    \multicolumn{1}{c}{$v_{a\varphi}$} & 
    \multicolumn{1}{c}{$f_Y$} & \multicolumn{1}{c}{$\tau$} &
    $\tilde{B}_r$&$\tilde{B}_m$
    \\[0.3ex]
    \hline
    \\[-2ex]
    $\la 0$& 0.8 &  15 &    6 & 100 & 0.1 & --12 & --13 \\
    0.6    & 0.3 &  36 &   15 &  30 & 8   & --10 & --16 \\
    1.2    & 0.8 &  38 &   17 &  30 & 8   & --12 & --16 \\[0.3ex]
    \hline
  \end{tabular*}\par \vspace{1ex}
  \footnotesize
  \hspace{1em} {\scshape Note.---}
  We list approximate values for the intrinsic color, distance,
  velocity dispersion, asymmetric drift, percentage of young-stars,
  approximate average age [Gyr], and the raw and mode-mixing
  corrected Oort's $B$ ($\tilde{B}_r$ and $\tilde{B}_m$, respectively).
\end{table}

%
%
%


In fact this is the only relevant property in which they differ
significantly from the red giants, see \Tab{sub_pops} and
\Fig{Oort.vs.BV}, in particular if one considers the mode-mixing
corrected values. A straightforward interpretation is that differences
in sample depth are unimportant in causing the differences in the
observed Oort constants. This in turn implies that small-scale wiggles
of the Galactic velocity field cannot be important either, for
otherwise we would expect significantly different results for red MS
stars and giants.


We can use the asymmetric drift as a proxy for the mean age and
velocity dispersion of a stellar population\footnote{Recall that stars
  obey an age-velocity dispersion relation \citep[$\sigma_*^2 \propto
  t$; e.g.,][]{J92} as well as Str\"omberg's asymmetric drift relation
  ($\sigma^2_* \propto v_{a\varphi}$, e.g.\ DB98), i.e.\ $v_{a\varphi}
  \propto t\propto \sigma^2_*$.}.  Similarly, we may use the
overabundance $\delta n(\ell)$ of stars in directions of spiral-arm
tangents as a proxy for the fraction of young stars.  In
\Fig{Oort.vs.Vad.Dnl}, we plot the Oort constants measured, including
the $m=3$ Fourier coefficients, against our estimate \eq{pi-w} for the
asymmetric drift ({\em left\/}) and $\delta n(\ell)$ ({\em right\/}).
The raw data (circles) show only marginal trends with $v_{a\varphi}$,
and perhaps even some dichotomy between red main-sequence (small
symbols) and giant populations (large symbols).  On the other hand,
the mode-mixing solutions do not show such a dichotomy but rather
tight linear relations between the Oort constants and the asymmetric
drift.

A dichotomy is clearly present in the plots versus $\delta n(\ell)$,
which arises because the red-clump stars at $\delta n(\ell)
\approx0.13$ have values similar to those of the red giants and unlike
those of the blue MS stars at the same $\delta n(\ell)$. Remarkable,
however, is that the red MS stars and red giants at $\delta
n(\ell)\approx0.04$ show rather similar values, despite that fact that
their average distances differ by a factor $>2$. This is true even for
the high-order terms, which argues for their non-zero values not being
created by sample-depth effects.

In \Sec{adrift} we saw that there should be almost no correlation
between asymmetric drift and Oort's $B$, for the case of axisymmetric
equilibrium (and constant sample depth). Thus, the very presence of
such a correlation with $v_{a,\varphi}$ as well as non-zero $C$
and $\tilde{s}_{\ell,3}$, argue strongly for non-axisymmetry to be the
dominant origin of the observed differences of the Oort constants
between stellar subsamples.

Alternatively, non-equilibrium effects may play a role. However, one
would then expect a somewhat erratic behaviour, instead of the clean
trends seen in the left plot of \Fig{Oort.vs.Vad.Dnl} and also in
\Fig{Oort.vs.BV} for the main-sequence stars (i.e. from $(\bv)_0=-0.3$
to 0.6).

To summarize, the variations of the Oort constants between the
subsamples originate most likely in deviations from axisymmetry, while
non-equilibrium effects and small-scale wiggles (deviation from
smoothness) seem less important (as do any effects that rely on
variations in sample depth). This in turn implies that any
interpretation of the Oort constants in terms of the properties of the
underlying Galactic potential, the very motivation to undertake
studies like this, see \Sec{intro}, cannot be as simple as in Oort's
days.
\section{Summary and Conclusion} \label{sec:sum}
the Oort constants are defined as the divergence ($K$), vorticity ($B$),
azimuthal ($A$) and radial ($C$) shear of the local stellar streaming
field of the Milky Way in the limit of vanishing random motions, where
all stars move on closed orbits, i.e.\ circular orbits in case of
axisymmetry.  The importance of the Oort constants derives from the fact
that the dynamics of these closed orbits is directly related to the
Galactic gravitational potential, a relation that becomes particularly
simple in the axisymmetric case.

The longitudinal proper motion of a star with parallax $\pi$ and
radial and azimuthal velocity with respect to the Sun, $U$ and $V$,
may be written as
\begin{equation} \label{eq:mu-ell}
  \o{\mu}_{\ell^\star} = 
  - {\rm Re}\,\big\{\pi (V+i\,U) \mathrm{e}^{i\ell}\big\}.
\end{equation}
Assuming that the stellar kinematics and parallaxes are uncorrelated,
one finds for the {\em mean\/} longitudinal proper motion
\begin{equation} \label{eq:mean-mu-ell}
  \o{\mu}_{\ell^\star} =
  - {\rm Re}\,\big\{\o{\pi} (\o{V}+i\,\o{U}) \mathrm{e}^{i\ell}\big\}.
\end{equation}
The spatial variations of $\o{U}$ and $\o{V}$ are given by the Oort
constants, and lead to a $|m|=1$ harmonics $\mathrm{e}^{im\ell}$ in
their Fourier expansion.  Together with the $\mathrm{e}^{i\ell}$ in
\eq{mean-mu-ell} this results in $m=0, \,2$ harmonics with amplitudes
given by the Oort constants $A$, $B$, and $C$. A similar harmonic
dependence is exhibited by the mean radial velocity times parallax,
$\o{\pi v_r}$. However, stellar parallaxes and radial velocities are
difficult to measure and thus the Oort constants are most commonly
determined from their effect on the stellar proper motions.

\begin{deluxetable}{l@{\quad}l@{\quad}p{13cm}}
   \ifpreprint
   \else
   \tabletypesize{\footnotesize}
   \fi
   \tablewidth{0pt}
   \tablecaption{The Oort Constants: true and measurable\label{tab:oort}}
   \tablehead{
                                  &
                \colhead{symbols} & 
                \colhead{definition and description}
   }
   \startdata
      (1)                        & 
          $A$, $B$, $C$, $K$         & the Oort constants:\\
                                 &   & divergence, vorticity, and shear of 
                                       the (hypothetical) velocity field 
                                       $\B{v}(\B{x})$ due to the closed orbits 
                                       supported by the Galactic potential 
                                     \\[.5ex]
      (2) \tablenotemark{a}      &
          $\o{A}$, $\o{B}$, $\o{C}$, $\o{K}$ & the best we can hope to get:\\
                                 &   & divergence, vorticity, and shear of 
                                       the streaming velocity field 
                                       $\o{\B{v}}(\B{x})$ of a group of stars 
                                     \\[.5ex]
      (3) \tablenotemark{b}     &
          $\tilde{A}$, $\tilde{B}$, $\tilde{C}$, $\tilde{K}$ &
          what we can actually measure:\\
                                &    & Fourier coefficients of the proper 
                                       motions measured for a group of stars 
   \enddata
   \tablenotetext{a}{
     Possible reasons for differences between (1) and (2):
     \newcounter{lll}
     \begin{list}{(\roman{lll})}
       {\usecounter{lll}\leftmargin4em\labelwidth3em\labelsep0.5em
         \itemsep0mm\parsep0.5ex\topsep0mm} 
     \item   For young stars: moving groups and other 
       non-equilibrium effects like spiral arms lead
       to unpredictable deviations of $\o{\B{v}}$ from a
       closed-orbit streaming field.
     \item   Local anomalies in the streaming field are reflected
       in the Oort constants if the sampling volume is
       too small. This is mainly affects
       sub-populations with low velocity dispersion
       making them susceptible to small-scale
       variations in the Galactic force field.
     \item   For old stars: $\o{\B{v}}$ deviates from closed-orbit 
       streaming by the asymmetric drift. For the axisymmetric
       case, this effect can be estimated using Str\"omberg's 
       asymmetric-drift relations: it reduces $A$ by up to 
       $3\kmskpc$, but hardly changes $B$.
     \end{list}
   }
   \tablenotetext{b}{
     Possible reasons for differences between (2) and (3):
     \begin{list}{(\roman{lll})}
       {\usecounter{lll}\setcounter{lll}{3}\leftmargin4em\labelwidth3em
         \labelsep0.5em\itemsep0mm\parsep0.5ex\topsep0mm} 
     \item   Correlations between the stellar parallaxes and velocities,
       which may occur for stars associated with spiral arms or a local
       warp, invalidate the basic assumption underlying the Fourier approach.
     \item   Terms of higher order than linear (= the Oort constants)
       in the Taylor expansion of the streaming field
       become increasingly important with ever deeper
       samples.
     \item   Discontinuities of the streaming field at the OLR of 
       the Galactic bar render the Oort constants
       ill-defined. If the sampling volume contains
       such places of resonance, the volume-averaged
       streaming field deviates systematically from the
       local field.
     \item   Mode mixing: variations of $\o{\pi}$ with $\ell$ in
       conjunction with the solar reflex motion lead to
       contributions to the proper motion that are
       indistinguishable from the Oort constants and up
       to a few \kmskpc\ in size.
     \end{list}
     }
\end{deluxetable}

\subsection{Mode Mixing and Other Problems}
The effect of the Oort constants on the stellar proper motions is
comparably small. For nearby stars (within ${\sim}1\,\kpc$), it is
much smaller than the contributions from random stellar motions (i.e.\ 
dispersion of $U$ and $V$ in \eqn{mu-ell} and the reflex of the solar
motion (i.e.\ the lowest order in \eqn{mean-mu-ell}. Therefore, large
proper motion surveys are necessary to extract the Oort constants with
reasonable accuracy.  There are various, mostly known but neglected,
problems arising in this procedure, which are summarized in the
footnotes of \Tab{oort}.

A fundamental, hitherto apparently overlooked, problem in measuring
the Oort constants from proper motion data is what we call {\em mode
  mixing} (point vii in \Tab{oort}). In fact, the cause of the problem
is rather similar to the very effect one is after. As a spatial
variation of $(\o{U},\o{V})$, described by the Oort constants,
contributes to the $m=0,\,2$ harmonics in the Fourier expansion of
$\o{\mu}_{\ell^\star}$, so does a variation of $\o{\pi}$. This
contribution is indistinguishable from that due to the Oort constants
itself.

For a typical situation, already a variation in the mean parallax of
only 10\% results in a contribution of a few \kmskpc, larger than any
other source of uncertainty in the Oort constants. Such a variation of
$\o{\pi}$ is not anticipated from a smooth exponential disk. However,
that seems to be a bad description of the actual situation for a
typical stellar sample. First, the stellar distribution is
inhomogeneous, in particular for early-type stars, which are
predominantly situated in spiral arms. Secondly, even if the
underlying density is rather smooth, extinction inevitable leads to
significant inhomogeneities in any actual stellar sample.

In order to correct for the mode mixing, one needs an unbiased
estimate of the variations of $\o{\pi}$ with $\ell$. For low-latitude
samples, such an estimate may be provided from the latitudinal proper
motion $\mu_b$, the mean of which at $b=0$ should be given by
$\o{\mu}_b=\o{\pi\,W}$. Here, $W$ is the vertical stellar motion with
respect to the Sun. If (1) $\pi$ and $W$ are uncorrelated, and (2)
$\o{W}$ is constant with $\ell$, then we can directly measure the
relative variations of $\o{\pi}$. A local stellar warp would
invalidate both these assumptions, and, hence, one would best not rely
on them, but use an independent and unbiased, though not necessarily
very accurate, distance estimate for the individual stars.
Unfortunately, such an estimate is not currently available -- note
that the photometric parallax is ill-suited for this purpose.

There is also a more technical issue that we want to emphasize. The
standard way to determine the Oort constant from proper motion surveys
used to be a least-square fit of the data to a functional form that
just accounts for the reflex of the solar motion and the Oort constants,
i.e.\ does not allow for higher-order Fourier modes. This essentially
is a parametric fit, and as such will give biased estimates for the
parameters if the data are not well described by the functional form
fitted. In our analysis, we allow for high-order terms, and found
indeed that these are non-negligible. We also restricted our analysis
to low latitude stars for three reasons: (i) restricting the latitude
to a narrow range, we do not need to account for possible latitudinal
variations of the mean proper motions; (ii) only at low $b$ can we use
the latitudinal proper motion to correct for mode mixing; and (iii)
the Oort constants are strictly defined only for closed, i.e.\ planar,
orbits. Moreover, most disk stars in sufficiently deep samples are
predominantly at low latitudes anyway.  A disadvantage of low latitude
stars is that extinction is largest and most patchy in this region, so
that the variation of $\o{\pi}$ with longitude is likely to be maximal
here.

\subsection{Longitudinal Number-Density Variations}
Before analyzing the proper motions of the low-latitude
($|b|<5.73\arcdeg$) stars from the ACT/Tycho-2 catalogs, we first
considered their longitudinal distribution as a function of color. For
the bluest stars in the sample ($\bv \approx{-}0.1$), this
distribution shows three narrow peaks in the direction of the nearby
spiral-arm tangents.

Next, we quantified the over-abundance of stars at the longitudes
associated with these three peaks. This over-abundance may be
interpreted as a measure for the importance of stars in spiral arms.
In fact, we use the relative over-abundance $\delta n(\ell)$ to
estimate the fraction of young stars in our color bins.  These
estimates compare well with estimates based on the star-formation
history.  With increasing color along the main sequence, $\delta
n(\ell)$ decreases, but never diminishes completely.

At colors redder than $(\bv)_0\approx 0.75$, sub-giant, red-clump,
giant and super-giant stars dominate the sample. Interestingly, at
about $(\bv)_0=0.9$, i.e.\ for red-clump stars, the over-abundance of
stars at the peak-longitudes increases back to the value for
early-type stars as blue as $(\bv)_0\approx 0.25$. This can be
explained by the about equal duration of the horizontal-branch phase
for stars of all masses ${\la}2 \, {\rm M}_\odot$, making the red
clump a strange mixture of stars of all ages. As a consequence, any
external observer of the Milky Way using some red pass band which
allows for red-clump stars to contribute significantly, would see a
spiral pattern considerably stronger than that of the underlying
old-disk population, which essentially represents the stellar mass
distribution.

For yet redder colors, i.e.\ beyond $(\bv)_0\approx1.2$, our measure
for the importance of spiral structure drops again, but not quite as
low as for the dwarfs at $(\bv)_0\approx0.6$. This is because the
duration of the giant-branch phase is shorter for more massive and
hence younger stars, such that the giants are rather old on average.

\subsection{The Oort Constants} 
\label{sec:Summary_OC}
The significant non-uniformity of the longitudinal stellar
distribution strongly hints towards inhomogeneity in the spatial
distribution of the sample stars. Thus, longitudinal parallax
variations and hence a confusion of the Oort constants by mode mixing
with the solar reflex motion is expected {\em and} observed.  In the
previous sections we have presented evidence that the mode-mixing
corrected values are likely to be close to the ``true'' Oort
constants.  We thus present the mode-mixing results below.

The absolute values of the Oort constants are small for young, blue
stars. Extrapolating to zero asymmetric drift we find (in \kmskpc).
\begin{equation} \label{eq:results_young}
  \mbox{young MS stars:\hspace{2em}}
  \begin{array}{r@{\,=\,}r}
    A          &   9.6,   \\ 
    B          & -11.6,   \\ 
    A-B        &  21.1,   \\ 
    C          &   0.4,   \\
    s_{\ell,3} &  -6.7,   \\
    c_{\ell,3} &  -0.1,   \end{array} 
\end{equation}
with internal errors of about 0.5\,\kmskpc. The oldest stars in our
samples, giants with $(\bv)_0\ga 1.2$, have significantly larger Oort
constants. At $v_{a\varphi}=18$ we get:
\begin{equation} \label{eq:results_old}
  \mbox{old red giants:\hspace{2em}}
  \begin{array}{r@{\,=\,}r}
    A          &  15.9 ,   \\ 
    B          & -16.9 ,   \\ 
    A-B        &  32.8 ,   \\ 
    C          &  -9.8 ,   \\
    s_{\ell,3} &   2.6 ,   \\
    c_{\ell,3} &   4.3 ,   \end{array} 
\end{equation}
with internal errors of about 1.2\,\kmskpc, and where 2\,\kmskpc has
already been added to the $A$ value of the giants in order to correct
for the reduction by the asymmetric drift (cf. \Sec{adrift}). The
external errors are harder to estimate\footnote{If we assume that the
difference between the raw and mode-mixing corrected values are
indicative of such external errors, we find: [$ \delta A, \delta
s_{\ell, 3}, \delta B,$ $\delta \mbox{A-B}, \delta c_{\ell,3}, \delta
C] \approx [0.1, 0.5, 4.4, 4.3,4.3, 9.8] \approx$ [0.6, 19, 26, 13, 100,
100]\% for old stars, and almost zero for young stars.}.

Our values for $A$ and $B$ for the old stars are somewhat unusual,
which is largely due to our mode-mixing correction.  Using the raw
Fourier coefficients whereby ignoring this effect, one finds for the
giants more conventional numbers, in particular
$A_{\mathrm{raw}}\approx14, B_{\mathrm{raw}}\approx{-}12.5$ and
$C_{\mathrm{raw}}\approx0$.

The values for the Oort constants for the young and old stars derived
{\em with} the mode-mixing corrections are not inconsistent with
previous determinations.  In part this arises due to the fact that
most previous work has applied parametric methods to derive the Oort
constants, in part because published results still have significant
uncertainties, and partly because all these values are uncertain due
to the various systematic effects discussed above.

A relatively large value for $\Omega$ has been found from the
longitudinal proper motion of Sgr~A$\!^\star$ (if Sgr~A$\!^\star$ is
at rest in the Galactic center). In that case, $\Omega\equiv A-B=
28\pm2$ \citep{BS99,Rea99}. Even larger values for $A-B (\approx
31.5$) have been reported by \citet{Mea99} and \citet{MZ98}, based on
absolute proper motions derived from the Southern Proper Motion
Program and the Hipparcos Catalogue, respectively (note, however, hat
these two works have employed the classical parametric $\mu_\ell = B +
A \cos2\ell$ description of Galactic rotation). All of these results
are consistent with our determinations for old stars \eq{results_old}.

\subsection{What can be learned from the Oort constants?}
To our knowledge, the linear dependence of the Oort constants on
asymmetric drift is a new discovery. It is worth noting here, that while
such a dependence was in principal expected already by Oort, our
measurements are inconsistent with expectations from axisymmetric models
for the Milky Way.

The non-zero $\tilde{s}_{\ell,3}$ value for young stars clearly shows
that non-axisymmetry is present amongst these stars, which given their
affiliation with spiral arms is not surprising at all. While for older
stars $\tilde{s}_{\ell,3}$ diminishes, another indicator of
non-axisymmetry, Oort's $C$, becomes very significantly non-zero.
For other sitations than axisymmetry, predictions for the values
expected for the Oort constants are considerably involved and require
detailled modelling. This is true especcially, if effects of the sample
depth and velocity distribution are to be taken into account.

As indicated by the imprints of the Galactic bar on the local velocity
distribution measured from Hipparcos data \citep{d98}, the Sun lies
just outside the outer Lindblad resonance (OLR) of the bar
\citep{d00}. This is exactly the place where the bar is expected to
most strongly distort the local kinematics. In particular, stars from
inside the OLR should produce $C<0$, and in a deep sample the number
of such stars should be larger than in a very local volume. Indeed,
when analyzing only the stars brighter than $V_T=10.5$, $C$ rises to
$-7.5$, i.e.\ the effect is weaker. Thus, the value of $C$ might arise
from contributions of stars from inside the OLR. \citet{MD03} have
analysed the effect of the Galactic bar on the stellar kinematics in
the outer disk and find that larger a velocity dispersion, as for the
red giants, tends to shift the ``effective radius'' of resonance
outwards. They also find weaker distortions to the {\em local\/} $C$
than our result.

Another intriguing possibility is that asymmetries in the stellar
distribution function arising from a (or more) satellite merger, like
that of the Sgr dwarf, contributes to the apparent anomalies. Note
that there appear to be asymmetries in the local number of thick
disk/inner halo stars \citep{lh96,phl01}.

Further work is needed to investigate whether such values can be
explained when (1) considering sample-depth effects, i.e.\ simulating
the actual data-taking and measurement process, (2) residual
non-equilibrium effects from minor mergers, (3) and accounting for the
additional non-axisymmetric forcing of triaxial halo, elliptical disk
\citep[e.g.,][]{kt94}, or spiral arms. This latter option is
interesting as the Sun resides also close to the co-rotation radius of
the spiral pattern \citep[e.g.,][]{LYS69,MPS79}, where one would
expect its forcing to be most efficient.
\subsection{Resum\'e}
The very concept of the Oort constants is originally based on the
idealization of the Milky Way being axisymmetric with almost circular
stellar orbits.  Similarly, in determining the Oort constants from
stellar proper motions one has usually relied, at least implicitly, on
the smoothness of the spatial distribution of the sampled stars. At
Oort's time, the deviations of reality from these ideals have been of
second order compared to the internal uncertainties due to the low
accuracy and sparse samples. However, with ever increasing accuracy
and larger samples, these deviations inevitably become significant.
Nonetheless, they have been largely ignored so far, resulting in
significant systematic errors. The presence of such errors is
reflected in the previous determinations of the Oort constants listed
by \citet{klb86}, whose scatter exceeds the typical internal errors.

In this paper, we (1) tried to explain the possible deviations from
the above ideal and the resulting systematics, in particular when
using large proper motion surveys, and (2) tried as far as possible
both to avoid and correct for them in our application to the
ACT/Tycho-2 catalogs. Avoiding uncontrollable systematics lead us to
exclude the results from either young, high-latitude, or nearby stars.
Thus, the most reliable results reported in the present study are
those for the (mainly old) giants in the Galactic plane, for which
there are two main sources of systematics: the asymmetric drift and
mode mixing. While the first problem was already known to
\citet{oort28}, the second is not mentioned in previous determinations
of the Oort constants from proper motions. Our correction for this
latter effect using the stars' vertical proper motions changes the
Oort $A$, $B$, and $C$ from more conventional numbers to the values
reported in \eqn{results_old} for old stars.

We hope that possible future space missions such as 
\ifpdf
\href{http://www.usno.navy.mil/FAME/}{FAME},
\href{http://www.aip.de/groups/DIVA/}{DIVA},
\href{http://sim.jpl.nasa.gov/}{SIM}, and
\href{http://astro.estec.esa.nl/GAIA/gaia.html}{GAIA}%
\else
FAME, DIVA, SIM and GAIA%
\fi
\footnote{
{\bfseries FAME:}http://www.usno.navy.mil/FAME/  \\
{\bfseries DIVA:}~http://www.aip.de/groups/DIVA/  \\
{\bfseries SIM:}~~~http://sim.jpl.nasa.gov/        \\
{\bfseries GAIA:}~http://astro.estec.esa.nl/GAIA/gaia.html
}
will dramatically improve both the quality and amount of the data,
yielding highly accurate five- or even six-dimensional phase-space
coordinates for millions of stars throughout the Milky Way. With such
data, many of the hitherto unpleasant deviations from Oort's idealized
Galaxy will become clearly apparent and must be dealt with properly.
Very likely, the classical deductive way of `measuring' the Oort
constants directly from the data, becomes hardly viable then.
Instead, a detailed analysis is needed which compares the data with
predictions from sophisticated dynamical models incorporating the
Galactic bar, spiral structure, non-equilibrium effects such as moving
groups, and other deviations from Oort's ideal Milky Way.  In fact,
these deviations, which plague our current determination of the
Galactic potential, are likely to constrain the models significantly
leading eventually to a much enhanced understanding of the structure
and formation of our host galaxy.

One datum that is very important to the study of the dynamics of the
Milky Way will not be collected by the future space-based astrometric
missions.  The lack of approximate ages of the
individual\footnote{Ages {\em can} be determined rather well for the
subsample of detached eclipsing binaries (approximately 3\% of
stars).}  tracer stars will seriously compromise the study of the
dynamical evolution of the Galaxy.  We have already noted that the
interpretation of the proper motions of certain kinds of stars (the
red-clump region) is hampered due our inability to associate these
stars with either young or old stellar populations.

We anticipate that age-related effects will become an important
limitation to the interpretation of the space-based astrometric data.
Age information would allow us to study many topics related to the
formation and evolution of the Milky Way, such as (1) the (relative)
ages of various stellar components, (2) the temporal variation of the
star formation activity, (3) the minor-merger history (through
distinct features in the age-velocity relation). Such studies would
turn the Milky Way into an important benchmark galaxy, not only at the
present, but also at cosmologically interesting epochs.

\acknowledgments
Substantial parts of this work were performed at Oxford, Heildeberg
and Potsdam (WD), and Southampton, Rutgers and USNO (RPO).

%
%

\ifpreprint
  \def\thebibliography#1{\section*{REFERENCES}
    \list{\null}{\leftmargin 1.2em\labelwidth0pt\labelsep0pt\itemindent -1.2em
    \itemsep0pt plus 0.1pt
    \parsep0pt plus 0.1pt
    \parskip0pt plus 0.1pt
    \usecounter{enumi}}
    \def\refpar{\relax}
    \def\newblock{\hskip .11em plus .33em minus .07em}
    \sloppy\clubpenalty4000\widowpenalty4000
    \sfcode`\.=1000\relax}
\else 
\fi

\end{document}